\documentclass[twocolumn,eqsecnum,aps]{revtex4}
\usepackage{epsfig}

\def\rhovec{\mbox{\boldmath $\rho$}}

\newcommand{\beq}{\begin{eqnarray}}
\newcommand{\eeq}{\end{eqnarray}}

\begin{document}

%\parindent=10pt
%%%%%%%%%%%%%%%%%%%%%%%%%%%%%%%%%%%%%%%%%%%%%%%%%%%%%%%%%%%%%%%%

\title {Light $\Xi$ hypernuclei in four-body cluster models}

\author{E.\ Hiyama}

\address{Nishina Center for
Accelerator-Based Science,
Institute for Physical and Chemical
Research (RIKEN), Wako, Saitama,
351-0198,Japan}

\author{Y. Yamamoto}

\address{Physics Section, Tsuru University, Tsuru, Yamanashi 402-8555, Japan}

\author{T. Motoba}

\address{Laboratory of Physics, Osaka Electro-Comm.
University, Neyagawa 572-8530, Japan}

\author{Th.~A. Rijken}

\address{Institute for Theoretical Physics,
University of Nijmegen, Nijmegen, The Netherlands}

\author{M. Kamimura}

\address{Department of Physics, Kyushu University,
812-8581,Japan}

%\date{\today}
%
%%%%%%%%%%%%%%%%%%%%%%%%%%%%


\begin{abstract}
Detailed structure calculations in
$^{\: 12}_{\Xi^-}$Be, $^{\ \: 5}_{\Xi^-}$H,
$^{\ \: 9}_{\Xi^-}$Li, $^{\ \: 7}_{\Xi^-}$H and
$^{\:10}_{\Xi^-}$Li are performed
within the framework of the microscopic
two-, three- and four-body cluster
models using the
Gaussian Expansion Method.
We adopted effective $\Xi N$ interactions
derived from the Nijmegen interaction models, which
give rise to substantially attractive $\Xi$-nucleus
potentials in accordance with the
experimental indications.
$^{\ \: 7}_{\Xi^-}$H and $^{\: 10}_{\Xi^-}$Li are
predicted to have bound states.
we propose to observe the
bound states in 
future $(K^-,K^+)$ experiments using
$^{\ \: 7}$Li and $^{\: 10}$B targets in addition to the
standard $^{12}$C target.
The experimental confirmation of these states
will provide   information on
the spin- and isospin-averaged $\Xi N$
interaction.

\end{abstract}

\maketitle

%===============================================================
\section{Introduction}
%===============================================================
%A standard procedure in hypernuclear physics is to investigate
%baryon-baryon interactions in strangeness $S<0$ sectors,
%whose features appear in many-body structures.
In studies of nuclear interactions, two-body scattering 
data are the primary input for characterizing interaction models.
However, $S=-1$ hyperon (Y)-nucleon (N) scattering data 
are very limited because of experimental issues.
For $S=-2$ interactions such as $\Lambda \Lambda$
and $\Xi N$, there are currently no scattering data.
Therefore, the existing $YN$ and $YY$ 
interaction models have a substantial degree of ambiguity.
Some $YN$ scattering experiments will be performed 
at the Japan Proton Accelerator Research Complex (J-PARC)
in the near future. Even at this facility, however, the 
possibility of performing $\Xi N$ or $\Lambda \Lambda$ 
scattering experiments is very limited or practically impossible.
Hence, in order to obtain useful information on $S=-2$
interactions, studies of many-body, hypernuclear structure 
are indispensable.

Our intention in this work is to investigate the possible existence 
of $\Xi$ hypernuclei and to explore the properties of the underlying 
$\Xi N$ interactions.  Identification of $\Xi$ hypernuclei in coming  
experiments at J-PARC will contribute significantly to understanding 
nuclear structure and interactions in $S=-2$ systems, which can
lead to an entrance into the world of multi-strangeness.
In order to encourage new experiments seeking $\Xi$ hypernuclei,
it is essential to make a detailed theoretical investigation of
the possible existence of  bound states, despite some uncertainty
in contemporary $\Xi N$ interaction models.

We investigate here the binding energies and 
structure of $\Xi$ hypernuclei produced by
$(K^-,K^+)$ reactions on light targets
on the basis of microscopic cluster models.
One of the primary issues is how to choose the $\Xi N$ interaction.
Although there are no definitive data for any $\Xi$ hypernucleus
at present, a few experimental data indicate that 
$\Xi$-nucleus interactions are attractive.
One example is the observed spectrum of the $(K^-,K^+)$ reaction
on a $^{12}$C target, where the cross sections for $\Xi^-$
production in the threshold region can be interpreted
by assuming a $\Xi$-nucleus Wood-Saxon (WS) potential
with a depth of $\sim$ $14$ MeV~\cite{E885}. 
Other indications 
of attractive $\Xi$-nucleus interactions are given by
certain emulsion data, the events for twin-$\Lambda$ hypernuclei,
where the initial $\Xi^-$ energies were determined by
the identification of all fragments after the
$\Xi^- p$-$\Lambda \Lambda$ conversion in nuclei.
The inferred $\Xi^-$ binding energies are substantially
larger than those obtained using only the Coulomb 
interaction~\cite{twin}.
When these $\Xi^-$ states are assumed to be  $1p$ states,
the WS potentials obtained from the binding energies are 
similar to the one above. 
These data suggest that the average $\Xi N$ interaction
should be attractive, which we utilize to select
the appropriate interaction models.
In this work we adopt  two types of $\Xi N$ interactions,
the Nijmegen Hard-Core model D (ND)~\cite{NHC-D}
and the Extended Soft-Core model (ESC04)~\cite{RY05,RY06}.

The structure of light $p$-shell nuclei can be reasonably
described in terms of cluster models composed of 
two- or three-body subunits. Here, we model the
possible $\Xi^-$ hypernuclei produced by $(K^-,K^+)$ 
reactions on available light $p$-shell targets as
four-body cluster structures:
The possible targets $^{12}$C, $^{11}$B, $^{10}$B, $^{9}$Be 
and $^{7}$Li naturally lead to such cluster configurations as
$\alpha\alpha t\Xi^-(^{\: 12}_{\Xi^-}$Be),
$\alpha \alpha 2n \Xi^-(^{\: 11}_{\Xi^-}$Li),
$\alpha \alpha n \Xi^-(^{\: 10}_{\Xi^-}$Li), 
$\alpha t n \Xi^-(^{\ \: 9}_{\Xi^-}$He) and
$\alpha nn\Xi^-(^{\ \: 7}_{\Xi^-}$H), respectively, by conversion of
a proton into a $\Xi^-$.  (In our model calculations, the
$\alpha \Xi^-$ potential is generated from a G-matrix $\Xi N$
interaction via a folding procedure.)  Here, among the above
$\Xi^-$ hypernuclei,
$^{\: \ 7}_{\Xi^-}$H($\alpha nn\Xi^-$) is the 
lightest $\Xi^-$ bound system, as shown in the following section.
In the case of lighter targets, $^6$Li, $^4$He, $^3$He and $d$,
the $\Xi^-$-hypernuclear states are composed of 
$\alpha n \Xi^-$, $pnn\Xi^-$ ($t\Xi^-$), 
$pn\Xi^-$ and $n\Xi^-$ configurations, 
respectively. However, these systems are not expected to support
bound states, considering the  weakly attractive nature of the
$\Xi N$ interactions suggested so far, except for Coulomb-bound 
(atomic) states.   Thus, possible $\Xi^-$ hypernuclear states to be 
investigated lie in the light $p$-shell region and may be considered
to have basically a four-body cluster structure.

This paper is organized as follows:
In Sec.II, we describe the basic properties of the $\Xi N$
interaction models and make clear what is relevant in the present
four-body calculations. In Sec.III, we perform the calculation
of $^{\: 12}_{\Xi^-}$Be$(\alpha \alpha t \Xi^-)$ with some
approximations, in order to fix the $\Xi N$ interaction strengths to
be consistent with the $(K^-,K^+)$ data.
The four-body cluster models, based on the 
Gaussian Expansion Method (GEM), have been
developed in a series of works for $\Lambda$ and
double-$\Lambda$ hypernuclei \cite{Hiyama96,
Hiyama97,Hiyama99,Hiyama00,Hiyama01,
Hiyama02,Hiyama06}.
In this work, similar cluster models are applied to
$^{\: 12}_{\Xi^-}$Be($\alpha \alpha t\Xi^-$),
$^{\:\ 7}_{\Xi^-}$H($\alpha nn\Xi^-$) and
$^{\: 10}_{\Xi^-}$Li($\alpha \alpha n\Xi^-$).
In Sec.IV, first we show the calculated behavior of the
$^{\: \ 5}_{\Xi^-}{\rm H}(\alpha \Xi^-)$ 
and $^{\: \ 9}_{\Xi^-}{\rm Li}(\alpha \alpha \Xi^-)$ systems, as a function
of the $k_F$ parameter in the $\Xi N$ G-matrix interaction,
to confirm the binding mechanism before adding neutron(s).
Then, in Sec.V we discuss the calculated results for
 $^{\: \ 7}_{\Xi^-}$H($\alpha nn\Xi^-$) and
$^{\: 10}_{\Xi^-}$Li($\alpha \alpha n\Xi^-$).

%===============================================================
\section{$\Xi N$ interactions}
%===============================================================

As stated above, the experimental information on 
$\Xi N$ interactions is quite uncertain.
It should be complemented by theoretical considerations.
Various $SU_3$-based interaction models have been 
proposed so far. 
In the construction of these models,
the scarce $YN$ scattering data are supplemented by the rich $NN$ 
scattering data through use of $SU_3$ relations among the 
meson-baryon coupling constants.
Though these models are more or less similar
in $S=-1$ systems, their $S=-2$ $\Xi N$ predictions differ 
dramatically from one another; most  are repulsive 
on average. In order to generate an attractive $\Xi N$ 
interaction on the basis of OBE modeling, it seems to 
be necessary that specific features are imposed.
In the past, the ND model has been popular for 
$S=-2$ interactions, because this model is compatible 
with the strong $\Lambda \Lambda$ attraction indicated by 
the older data on double $\Lambda$ hypernuclei, and also
it yields attractive $\Xi$-nucleus interactions.
These aspects of ND are the result of its specific feature
that the unitary-singlet scalar meson is included 
without any scalar-octet mesons.
In this case, the strong $\Xi N$ attraction originates
from this scalar-singlet meson which gives the same 
contributions in all $YN$ and $YY$ channels. 
In the case of other Nijmegen OBE models, the attractive 
contributions of the scalar-singlet mesons are substantially
cancelled by those of the scalar-octet mesons, and
their $\Xi N$ sectors are repulsive on average.
A different OBE modeling for attractive $\Xi N$ interactions
has been adopted in the Ehime model~\cite{Ehime}, 
where the insufficient
$\Xi N$ attraction given by scalar-nonet mesons is 
supplemented by adding another scalar-singlet meson $\sigma$
and the coupling constant $g_{\Xi \Xi \sigma}$ is adjusted 
so as to give reasonable $\Xi N$ attraction, independent
of the SU3-relations among coupling constants.
The two models, ND and Ehime, are essentially similar, in 
that substantial parts of the $\Xi N$
attraction result from the scalar-singlet mesons.
%In this work, we use only Ehime, because the hard-core
%nature of NHC-D is not suitable for our four-body 
%variational method.

More recently, new interaction models ESC04 (a,b,c,d)
have been introduced, models in which two-meson and meson-pair 
exchanges are taken into account, and in principle 
no {\it ad hoc} effective  boson-exchange potentials
are included~\cite{RY05,RY06}.
The features of the ESC04 models differ significantly from 
those of the OBE models, especially in the $S=-2$ channels.
Among the ESC04 models, ESC04d is distinguished,
because the resulting $\Xi$-nucleus interaction gives attraction
suggested by the above experimental situation.
%to what is experimentally indicated, leaving the strong
%attraction from the scalar mesons unchanged. 
%
This is mainly due to the following mechanism:
A remarkably strong attraction appears in the 
$T=0$ triplet-even ($^{13}S_1$) state, because the strongly 
repulsive contribution of vector mesons is cancelled 
by the attractive contributions from axial-vector mesons.
In fact, the attraction in this state is so strong that peculiar 
$\Xi$ bound states are produced in  few-body systems \cite{Yamamoto08},
though such considerations lie outside the scope of the 
work presented in this paper.
In later calculations, the important points are the spin- and 
isospin-averaged even-state interactions, which are strongly
attractive owing to the significant $^{13}S_1$-state attraction.
Another important feature of ESC04d in the $S=-2$ channel
is that the meson-pair exchange terms give rise to
strong $\Lambda \Lambda$-$\Xi N$-$\Sigma \Sigma$
and $\Xi N$-$\Lambda \Sigma$-$\Sigma \Sigma$ 
coupling interactions. This feature of ESC04d makes
the conversion widths of $\Xi$-hypernuclear states
far larger than those for ND.

 \begin{table}[h]
 \caption{Partial wave contributions to $U_\Xi(\rho_0)$.
In the case of ESC04d, the medium-induced repulsion is
included by taking $\alpha_V=0.18$. In the case of ND,
the hard-core radii are taken as $r_c=0.52$ and 0.45 fm in
the $^{11}S_0$ and the other states, respectively. 
}
 \label{Gmat1}
% \begin{center}
  \begin{tabular}{lcrrrrrrrr}
  \hline
 model & $T$ & $^1S_0$ & $^3S_1$ & $^1P_1$ & $^3P_0$ & $^3P_1$ & $^3P_2$ 
 & \enskip $U_\Xi$ & \enskip$\Gamma_\Xi$ \\
  \hline
 ESC04d&$0$& 6.3 & --18.4 &   1.2 &  1.5  & --1.3 & --1.9 &  & \\
       &$1$& 7.2 & --1.7   & --0.8 & --0.5 & --1.2 & --2.5 & \enskip --12.1
       & \enskip12.7 \\
  \hline
  ND   &$0$& --3.0 & --0.5  & --2.1 & --0.2 & --0.7 & --1.9 &  & \\
       &$1$& --4.1 & --4.2  & --3.0 & --0.0 & --3.1 & --6.5 & \enskip--29.5
        &\enskip 0.8 \\
  \hline
  \end{tabular}
% \end{center}
 \end{table}

Our cluster models are composed of cluster units
($\alpha$ and $t$), $n$ and $\Xi^-$, where
the $\alpha (t) \Xi^-$ interactions are obtained by
folding the $\Xi N$ G-matrix interactions into
the density of $\alpha$($t$). 
According to the method described in Refs.~\cite{RY05,RY06},
the $\Xi N$ G-matrix interactions are derived from
ESC04d and ND in nuclear matter, where the imaginary parts
arise from the energy-conserving transitions from $\Xi N$ 
to $\Lambda \Lambda$ channels in the nuclear medium. 
The resulting complex G-matrix interactions are represented
as $k_F$-dependent local potentials 
\begin{equation}
G^{(\pm)}_{TS}(r,k_F)= \sum^3_{i=1}\, (a_i+b_i k_F +c_i k_F^2)\,
\exp {(-r^2/\beta_i^2)} \ ,
\label{eq:yng}
\end{equation}
where $k_F$ is the Fermi momentum of nuclear matter.
The suffixes $(+)$ and $(-)$ specify even and odd, respectively.
In our applications to finite $\Xi$ systems, it is plausible to 
obtain the $k_F$ values from the average density in the respective 
systems. In a similar G-matrix approach to $\Lambda$ hypernuclei,
for instance, the $\Lambda N$ G-matrix interactions can be adopted
to reproduce the observed $\Lambda$ binding energy ($B_\Lambda$)
by choosing appropriate $k_F$ values. 
Such a procedure cannot be applied strictly
in the case of $\Xi$ hypernuclei, because there exist no definitive
experimental data. In this work, we are obliged to choose the $k_F$
values rather arbitrarily but within a reasonable range 
($0.8 \sim 1.2$ fm$^{-1}$ in light $p$-shell systems). 
Here, the experimental indication for the existence of
$^{\: 12}_{\Xi^-}$Be is used to adopt the $\Xi N$ G-matrix interactions,
although it is not so definite because an experimental $\Xi$
binding energy ($B_\Xi$)
could not be extracted.
As shown later, the adjustable parts included in ND and ESC04d are 
determined so that the $\Xi$ $s$-state energy  
in our model of 
$^{\: 12}_{\Xi^-}$Be has the value $-2.2$ MeV for an adequate value
of $k_{\rm F}$, 
being obtained from the $\Xi$-nucleus WS potential with 
the depth $-14$ MeV~\cite{E885} where
the Coulomb interaction is switched off.
%In the case of Ehime, this adaptation can be realized
%by changing the coupling constant $g_{\Xi \Xi \sigma}$.
In the case of ND, this constraint can be realized by choosing
the hard-core radius $r_c$: 
We take $r_c=0.52$ and $0.45$ fm in the $^{11}S_0$ state and the 
other states, respectively. The former choice is made so that the 
derived $^{11}S_0$ $\Lambda \Lambda$ G-matrix interaction reproduces
the $\Lambda \Lambda$ bond energy observed in the double-$\Lambda$ 
hypernucleus.
On the other hand,
the constraint in the case of ESC04d is enforced by changing 
the parameter $\alpha_V$ controlling the medium-induced 
repulsion~\cite{RY05}: We take $\alpha_V=0.18$. 
%In these cases, the used values of $k_F$ are 1.1 fm$^{-1}$ 
%and 1.03 fm$^{-1}$, respectively, for Ehime and ESC04d.
%
Hereafter, ESC04d with $\alpha_V=0.18$ is denoted as ESC
for simplicity.

In the Table \ref{Gmat1},
we show the partial-wave contributions
of the resulting $\Xi$ potential depth $U_\Xi$ in nuclear matter
at normal density $\rho_0$ ($k_F=1.35$ fm$^{-1}$).
The $U_\Xi$ values are found to be very different for
ESC and ND, because the odd-state contributions in
the former are far more attractive than those in the latter.
It is noted, however, the odd-state interactions play minor
roles in light systems considered in this work.
More important is that the spin- and isospin-dependence
differs significantly between ESC and ND.

The interaction parameters ($a_i$, $b_i$ and $c_i$) in our
G-matrix interactions (\ref{eq:yng}) are tabulated in 
Tables \ref{Gmat2} and \ref{Gmat3} for ESC and ND, respectively.
Hereafter, G-matrix interactions derived from ESC and ND
are denoted as $G_{ESC}$ and $G_{ND}$, respectively.

\begin{figure}
 \resizebox{0.48\textwidth}{!}{%
 \includegraphics{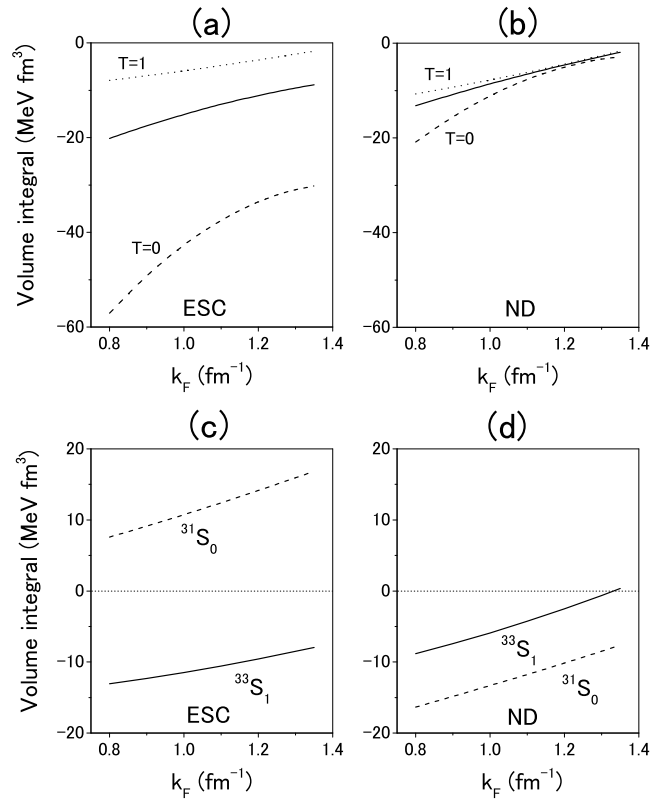}
 }
\caption{The volume integrals of $\bar G^{(+)}=
(G^{(+)}_{00}+3G^{(+)}_{01}+3G^{(+)}_{10}+9G^{(+)}_{11})/16$
are drawn as a function of $k_F$ by solid curves in (a) for ESC 
and in (b) for ND. Here, $T=0$ (dashed) and $T=1$ (dotted) 
parts show the volume integrals of $(G^{(+)}_{00}+3G^{(+)}_{01})/4$ 
and $(G^{(+)}_{10}+3G^{(+)}_{11})/4$, respectively.
In (c), the volume integrals of $^{33}S_1$ and $^{31}S_0$ 
components for ESC are drawn by solid and dashed curves, 
respectively. The corresponding ones for ND are in (d).
}
\label{Volint}
\end{figure}

The features of our G-matrix interactions can be demonstrated 
clearly by the volume integrals of the G-matrix interaction:
$J_V(k_F)=\int_0^\infty G(r,k_F) r^2 dr$. Here, we define
the spin- and isospin-averaged interactions as
$\bar G^{(\pm)}=
(G^{(\pm)}_{00}+3G^{(\pm)}_{01}+3G^{(\pm)}_{10}+9G^{(\pm)}_{11})/16$.
The volume integrals of $\bar G^{(+)}(r,k_F)$ are drawn
as a function of $k_F$ in Fig.~\ref{Volint}, where
(a) and (b) are for ESC and ND, respectively.
It should be noted here that the even-state interaction of
ESC is more attractive than that of ND.
In the cases of our cluster systems, $\Xi$-states are determined
dominantly by $\alpha$ $\Xi$ folding interactions derived from
$\bar G^{(\pm)}(r,k_F)$.
The $\bar G^{(-)}$ for ND is far more attractive than that for ESC, though
 their contributions in $s$-shell systems are very small.
Similarly, (c) and (d) in Fig.~\ref{Volint} show the volume 
integrals of the triplet- and singlet-even state interactions 
in the $T=1$ state for ESC and ND, respectively.
Here, the $^{33}S_1$ and $^{31}S_0$ interactions in ESC
are found to be attractive and repulsive, respectively.
On the other hand, both of $^{33}S_1$ and $^{31}S_0$ interactions
are attractive in ND, and the latter is more attractive than
the former. Namely, the $T=1$ spin-spin interaction in ND (ESC)
is repulsive (attractive).
This difference of the $T=1$ spin-spin interactions for ESC and ND
is reflected in the level structures of $^{\:\ 7}_{\Xi^-}$H
and $^{\: 10}_{\Xi^-}$Li,
as shown later. 

Another important difference  between ESC and ND is that the 
$\Lambda \Lambda$-$\Xi N$-$\Sigma \Sigma$ coupling interaction in the former
is far stronger than that in the latter.
This is reflected by the fact that the calculated value
of the conversion width $\Gamma_\Xi$ for ESC is far larger 
than that for ND, as exhibited in Table \ref{Gmat1}.

\begin{table}[h]
\caption{The parameters in the G-matrix interaction 
$G^{(\pm)}_{TS}(r,k_F)$ given by (\ref{eq:yng}) for ESC.
 Entries are given in units of $a$ [MeV], $b$ [MeV$\cdot$fm] 
 and $c$ [MeV$\cdot$fm$^2$] } 
\label{Gmat2} 
\vskip 0.3cm 
\begin{tabular}{ccccc}
\hline &  $\beta_i$ (fm) & 0.50 &  0.90  &  2.00  \\ \hline
         & $a$  &  0.0  & $-$690.8$-$309.0$i$ & $-$2.759 \\
$G^{(+)}_{00}$ & $b$  &  0.0  &    1263.+252.4$i$ &      0.0 \\
         & $c$  &  0.0  & $-$451.7$-$111.0$i$ &    0.0 \\
\hline
         & $a$  & $-$6959. &    756.5 & $-$1.317 \\
$G^{(+)}_{01}$ & $b$  &   11280. & $-$1567. &      0.0 \\
         & $c$  & $-$4371. &    627.2 &      0.0 \\
\hline
         & $a$  & $-$1634. &    257.8 & $-$1.528 \\
$G^{(-)}_{00}$ & $b$  &    3426. & $-$137.4 &      0.0 \\
         & $c$  & $-$965.8 &    60.78 &      0.0 \\
\hline
         & $a$  & $-$5692. &    175.0$-$15.50$i$ & $-$1.411 \\
$G^{(-)}_{01}$ & $b$  &    7697. & $-$583.9+24.31$i$ &  0.0 \\
         & $c$  & $-$2667. &    303.8$-$13.91$i$ &      0.0 \\
\hline
\hline
         & $a$  & $-$216.4 &    48.96 & $-$1.838 \\
$G^{(+)}_{10}$ & $b$  &    676.0 & $-$83.76 &      0.0 \\
         & $c$  & $-$198.1 &    43.36 &      0.0 \\
\hline
         & $a$  &    527.9 & $-$121.8 & $-$1.787 \\
$G^{(+)}_{11}$ & $b$  &    85.16 & $-$10.83 &      0.0 \\
         & $c$  &    13.25 &    9.351 &      0.0 \\
\hline
         & $a$  & $-$2671. &    36.08 & $-$1.043 \\
$G^{(-)}_{10}$ & $b$  &    3343. & $-$116.6 &      0.0 \\
         & $c$  & $-$1034. &    53.97 &      0.0 \\
\hline
         & $a$  &    1435. & $-$166.3 & $-$1.168 \\
$G^{(-)}_{11}$ & $b$  &    451.1 & $-$24.19 &      0.0 \\
         & $c$  & $-$131.2 &    18.44 &      0.0 \\
\hline
\end{tabular}
\end{table}

\begin{table}[h]
\caption{The parameters in the G-matrix interaction 
$G^{(\pm)}_{TS}(r,k_F)$ given by (\ref{eq:yng}) for ND. 
Entries are given in units of $a$ [MeV], $b$ [MeV$\cdot$fm] 
and $c$ [MeV$\cdot$fm$^2$] } 
\label{Gmat3}
\vskip 0.3cm 
\begin{tabular}{ccccc}\hline &  $\beta_i$ (fm) & 0.50 &  0.90  &  2.00  \\ \hline
         & $a$  & $-$8769. &    456.1$-$102.1$i$ & $-$2.505 \\
$G^{(+)}_{00}$ & $b$  &  15530.  & $-$1082.+91.45$i$ &  0.0 \\
         & $c$  & $-$6383. &    473.2$-$18.03$i$ &      0.0 \\
\hline
         & $a$  &    452.4 & $-$105.8 & $-$.6861 \\
$G^{(+)}_{01}$ & $b$  & $-$25.82 &    10.75 &      0.0 \\
         & $c$  &    67.40 &    10.22 &      0.0 \\
\hline
         & $a$  & $-$7382. & $-$168.9 & $-$3.141 \\
$G^{(-)}_{00}$ & $b$  &    8672. & $-$140.1 &      0.0 \\
         & $c$  & $-$3145. &    69.22 &      0.0 \\
\hline
         & $a$  & $-$569.3 & $-$231.1$-$7.788$i$ &    .0300 \\
$G^{(-)}_{01}$ & $b$  &    2072. & $-$32.47+6.124$i$ &  0.0 \\
         & $c$  & $-$696.8 &    22.21$-$.5631$i$ &      0.0 \\
\hline
\hline
         & $a$  &    356.9 & $-$138.5 & $-$.3949 \\
$G^{(+)}_{10}$ & $b$  &    110.3 &    13.97 &      0.0 \\
         & $c$  & $-$1.818 &    7.792 &      0.0 \\
\hline
         & $a$  &    436.0 & $-$108.4 & $-$1.334 \\
$G^{(+)}_{11}$ & $b$  &    6.513 &    10.11 &      0.0 \\
         & $c$  &    46.10 &    10.88 &      0.0 \\
\hline
         & $a$  &    75.12 & $-$254.3 &    .1086 \\
$G^{(-)}_{10}$ & $b$  &    939.6 & $-$.0260 &      0.0 \\
         & $c$  & $-$269.5 &    7.792 &      0.0 \\
\hline
         & $a$  & $-$281.0 & $-$218.5 & $-$1.003 \\
$G^{(-)}_{11}$ & $b$  &    1227. & $-$5.773 &      0.0 \\
         & $c$  & $-$422.7 &    11.82 &      0.0 \\
\hline
\end{tabular}
\end{table}

Our cluster models for $A=7$ and $10$ systems are composed of
$\alpha nn \Xi^-$
 and $\alpha \alpha n \Xi^-$ , respectively,
where the $\Xi N$ G-matrix interactions are used to obtain
$\alpha \Xi^-$ folding potentials based on the $(0s_{1/2})^4$
configuration with $b_N=1.358$ fm. It is problematic, 
on the other hand, to use the G-matrix interactions
for the $\Xi n$ parts. The reason is as follows: Correlations of
$\Xi n$ pairs are treated exactly in our model space spanned
by Gaussian functions, which means some double counting for
$\Xi n$ short-range correlations that has been already 
included in the G-matrix interactions. 
Though a reasonable way out of this problem is to use directly the bare 
potentials (ESC and ND), there appear some  difficulties in such treatments:
In the case of ESC, the $\Xi N$-$\Lambda \Sigma$ and $\Xi N$-$\Lambda 
\Sigma$-$\Sigma \Sigma$ coupling potentials in the $T=1$ channels make 
our treatment extremely complicated. In the case of ND, although these 
coupling potentials are not taken into account, the hard-core 
singularities cannot be treated in our Gaussian model space. 
Thus, we adopt 
here simple three-range Gaussian substitutes simulating the bare 
potentials. They are fitted so that the G-matrices derived from them 
simulate the original $T=1$ G-matrices at $k_F=1.0$ fm$^{-1}$. Here, 
the $\Xi N$-$\Lambda \Sigma$ and $\Xi N$-$\Lambda \Sigma$-$\Sigma 
\Sigma$ couplings in the ESC case are  effectively renormalized into 
the $\Xi N$ single-channel potentials.
The determined interaction parameters are given in Table \ref{Gmat4} 
for ESC and ND.

\begin{table}[ht]
\caption{Parameters of the three-range Gaussian interactions 
simulating (a) ESC and (b) ND in the $T=1$ $\Xi n$ states.
}
 \label{Gmat4}

\begin{tabular}{crrr}
&&& \\
(a) ESC &&& \\
\hline
$\beta_i$ (fm) & 0.40 & 0.80 & 1.50  \\
\hline
$^{31}E$ &  519.5 &    66.27 & $-$7.230 \\
$^{33}E$ &  217.4 & $-$170.0 & $-$7.058 \\
$^{31}O$ &    0.0 & $-$39.56 & $-$5.178 \\
$^{33}O$ &    0.0 & $-$55.40 & $-$6.936 \\
\hline

&&& \\
(b) ND &&& \\
\hline
$\beta_i$ (fm) & 0.50 & 0.90 & 2.00  \\
\hline
$^{31}E$ &  1076. & $-$159.6 & $-$5.432 \\ $^{33}E$ &  1331. & $-$134.0 & $-$7.610 \\
$^{31}O$ &    0.0 & $-$32.30 & $-$5.432 \\
$^{33}O$ &    0.0 &    18.16 & $-$7.610 \\
\hline

\end{tabular}

\end{table}

%===============================================================
\section{$^{\:12}_{\Xi^-}$B\lowercase{e}($\alpha \alpha t \Xi^-$) system}
%===============================================================

Let us start from the analysis for the
$^{\:\ 12}_{\Xi^-}$Be
($^{11}{\rm B}+\Xi^-$) hypernucleus  produced
by the $^{12}{\rm C}(K^-,K^+)$ reaction,
adopting the $\alpha \alpha t \Xi^-$ four-body model.
In this case,
the $(T,J^\pi) =(1,1^-)$ states are produced, 
because the $T_z$ component is transformed by 
$\Delta T_z=1$ on the $T=0$ target.
This system is important in a double sense.
One is that in  BNL-E885 a fairly deep $^{11}$B-$\Xi^-$ 
potential was indicated as mentioned in the previous section.
The other is that this reaction is planned as the Day-1 
experiment at J-PARC.

The above-mentioned G-matrix interactions $G_{ESC}$
and $G_{ND}$ are adjusted so as to be consistent with
the Woods-Saxon potential depth of
BNL-E885 
 within the
framework of the $\alpha \alpha t \Xi^-$ four-body model.

%===============================================================
\subsection{Model and Interaction}

\begin{figure}[htb]
\begin{center}
\epsfig{file=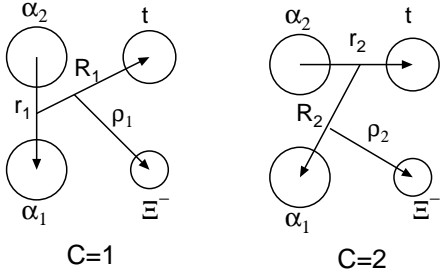,scale=0.5}
\end{center}
\caption{Jacobian coordinates 
for the $\alpha \alpha t \Xi^-(^{\: 12}_{\Xi^-}$Be)
four-body system. The two $\alpha$ clusters are to be
symmetrized.}
\label{fig:be12}
\end{figure}
In the case  of an $\alpha \alpha t \Xi^-$ four-body model,
we take two sets of  Jacobian coordinates as
shown in Fig.~\ref{fig:be12}, since we get sufficiently converged energies
using only those two sets of Jacobian coordinates.
The total Hamiltonian and the   
Schr\"{o}dinger equation
are given by 
\begin{equation}
 ( H - E ) \, \Psi_{JM}(^{\:12}_{\Xi^-}{\rm Be})  = 0 \ , \\
\label{eq:hamiltonian-12}
\end{equation}
\begin{equation}
 H=T+\sum_{a,b}V_{a b}
      +V_{\rm Pauli} \ ,
\label{eq:hamiltonian-12-1}
\end{equation}
where $T$ is the kinetic-energy operator and
$V_{ab}$ is the interaction between 
constituent particles $a$ and $b$.
The  OCM projection
operator $V_{\rm Pauli}$ will be given below.
The total wavefunction
is described as a sum of amplitudes
of the rearrangement channels $(c=1$ and 2)
of Fig.~\ref{fig:be12} in the $LS$ coupling
scheme:
\begin{eqnarray}
      \Psi_{JM,\:TT_z}\!\!&(&\!\!^{\: 12}_{\Xi^-}{\rm Be})
       =  \sum_{c=1}^{2}
      \sum_{n,N,\nu}  \sum_{l,L,\lambda}
       \sum_{S,I,K}
       C^{(c)}_{nlNL\nu\lambda S IK} \nonumber  \\
      &  \times & {\cal S}_\alpha
      \Big[
       \Phi(\alpha_1)\Phi(\alpha_2) 
     \big[ \Phi_{\frac{1}{2}}(t)
       \chi_{\frac{1}{2}}(\Xi^-)
       \big]_S   \nonumber \\
  & \times  &    \big[ \big[ \phi^{(c)}_{nl}({\bf r}_c)
         \psi^{(c)}_{NL}({\bf R}_c)\big]_I
        \xi^{(c)}_{\nu\lambda} (\mbox{\boldmath $\rho$}_c)
        \big]_{K}  \Big]_{JM} \nonumber \\
 & \times  &    \big[\eta_{\frac{1}{2}}(t) 
        \eta_{\frac{1}{2}}(\Xi^-) \big]_{T,T_z} \;  .
%(2.3)
\label{eq:wavefunction-12}
\end{eqnarray}
Here the operator ${\cal S}_\alpha$
stands for the symmetrization operator
for exchange of two $\alpha$ clusters.
$\chi_{\frac{1}{2}}(\Xi^-)$is the
spin function of the $\Xi^-$ particle and
$\eta_{\frac{1}{2}}(\Xi^-)$ is
the isospin function of the $\Xi^-$ particle.
Following the Gaussian Expansion Method (GEM)
\cite{Kami88,Kame89,Hiyama03},
we take the functional form of
$\phi_{nlm}({\bf r})$,
$\psi_{NLM}({\bf R})$ and
$\xi^{(c)}_{\nu\lambda\mu} (\mbox{\boldmath $\rho$}_c)$ as
\begin{eqnarray}
      \phi_{nlm}({\bf r})
      &=&
      r^l \, e^{-(r/r_n)^2}
       Y_{lm}({\widehat {\bf r}})  \;  ,
 \nonumber \\
      \psi_{NLM}({\bf R})
      &=&
       R^L \, e^{-(R/R_N)^2}
       Y_{LM}({\widehat {\bf R}})  \;  ,
 \nonumber \\
      \xi_{\nu\lambda\mu}(\mbox{\boldmath $\rho$})
      &=&
       \rho^\lambda \, e^{-(\rho/\rho_\nu)^2}
       Y_{\lambda\mu}({\widehat {\rhovec}})  \; ,
%(2.4)
\end{eqnarray}
where the Gaussian range parameters are chosen 
according to geometrical progressions:
\begin{eqnarray}
      r_n
      &=&
      r_1 a^{n-1} \qquad \enspace
      (n=1 - n_{\rm max}) \; ,
\nonumber\\
      R_N
      &=&
      R_1 A^{N-1} \quad
     (N \! =1 - N_{\rm max}) \; ,  %(2.5)
\nonumber\\
      \rho_\nu
      &=&
      \rho_1 \alpha^{\nu-1} \qquad
     (\nu \! =1 - \nu_{\rm max}) \; .  %(2.5)
%(2.5)
\end{eqnarray}
  The eigenenergy $E$  in Eq.(\ref{eq:hamiltonian-12})
and the coefficients $C$ in Eq.((\ref{eq:wavefunction-12})) 
are to be determined by the Rayleigh-Ritz variational method.

As for the $\alpha \alpha$ and $\alpha t$ interactions,
we employ the potentials which have been used often
in the OCM-based cluster-model study
of light nuclei:
Our potentials $V_{\alpha \alpha}$ \cite{Hasegawa71} 
and $V_{\alpha t}$\cite{Furutani80}
reproduce reasonably well the low-lying bound
states and low-energy scattering phase shifts of the
$\alpha \alpha$ and $\alpha t$ systems, respectively.
The Coulomb potentials are constructed by folding
the $p {\rm -} p$ Coulomb force into the proton
densities of all the 
participating  clusters.

The Pauli principle between nucleons belonging
to $\alpha$ and $x(=\alpha, t)$ clusters
is taken into account by the orthogonality condition
model (OCM) \cite{Saito69}.
The OCM projection operator $V_{\rm Pauli}$
appearing in Eq. (\ref{eq:hamiltonian-12-1})
is represented by
\begin{equation}
V_{\rm Pauli}=\lim_{\gamma\to\infty} \ \gamma \
\sum_f
|\phi_f({\bf r}_{\alpha x})
\rangle \langle \phi_f({\bf r}'_{\alpha x})| \:,
\label{eq:ocm}
\end {equation}
which rules out the amplitude of the
Pauli-forbidden $\alpha -x$ relative
states $\phi_f({\bf r}_{\alpha x})$
from the four-body total wavefunction \cite{Kukulin95}.
The forbidden states are
$f={0S,1S,0P,0D}$ for
$x=t$ and $f={0S,1S,0D}$ for $x=\alpha$.
The Gaussian range parameter $b$ of the
single-particle $0s$ orbit in the
$\alpha$ particle $(0s)^4$ is taken to be
$b=1.358$ fm so as to reproduce the
size of the $\alpha$ particle.
For simplicity the  same size is assumed for the $t$ cluster 
in  treating the Pauli principle.
In the actual calculations, the strength $\gamma$
for $V_{\rm Pauli}$ is taken to be
$10^4$ MeV, which is large enough to push
the unphysical forbidden state to
the very high energy region, while keeping the
physical states unchanged.

Using the $V_{\alpha \alpha}$ and $V_{\alpha t}$
potentials, we perform the three-body calculation for
the $^{11}{\rm B}(\alpha \alpha t)$ system.
The calculated values of the ground $(3/2^-_1)$ and
the first excited $(1/2^-_1$) states in $^{11}$B 
are overbound in comparison with the experimental values.
In order to put the subsequent four-body
calculations for $^{\:12}_{\Xi^-}$Be$(\alpha \alpha t \Xi^-)$
on a  sound basis,
we introduce a phenomenological
$\alpha \alpha t$ three-body force of the
following form:
\begin{equation}
V_{\alpha \alpha t}=v_0{\rm exp}[-(r_{\alpha_1 t}/\beta)^2
-(r_{\alpha_2 t}/\beta)^2] \: .
\end{equation}
Here we adopt $v_0=+95$MeV and $\beta=2.26$ fm
in order to reproduce the
$^{11}{\rm B}(3/2^-_1)$ ground state energy.
For the excitation energy of the
$^{11}{\rm B}(1/2^-_1)$ state,
we stick to  the exact experimental value instead of the
calculated value, when we perform
the hypernuclear 
four-body calculations.

%===============================================================
\subsection{Results for $^{\:12}_{\Xi^-}$Be and
the appropriate $k_{\rm F}$ parameter}

As mentioned before,
our $\Xi N$ interactions are adjusted so as to
give the $\Xi^-$ $s$-state energy $-2.2$ MeV
in the $^{\:12}_{\Xi^-}$Be system. 
This value is consistent
with the observed spectrum of the $^{12}{\rm C}(K^-,K^+)$ 
reaction which
suggests the WS potential
depth of 14 MeV \cite{E885}. 
If we assume spin-nonflip dominance for
 the $^{12}$C$(K^-,K^+)$ reaction,
 the $[p^{-1}_{3/2}\:s^{\Xi^-}_{1/2}]_{J=1^-}$
 state is naturally excited.
Therefore, within the framework of the
$\alpha \alpha t \Xi^-$ four-body model,
the $k_{\rm F}$ parameters in the
$\alpha \Xi^-$ and $t \Xi^-$ potentials, without Coulomb interaction,
are tuned so that
the $1^-_1$ state energies agree with $-2.2$ MeV.
\begin{table*}[htbp]
  \caption{The calculated $B_{\Xi^-}$(MeV) of
  the $1^-_1$ and $2^-_1$ states using ESC and ND potentials
  for $^{12}_{\Xi^-}$Be.
  In order to reproduce the 'observed' $B_{\Xi^-}$,
  we tuned $k_{\rm F}=1.055$(fm$^{-1}$) and 1.025(fm$^{-1}$) for
  the ESC and ND potentials, respectively.
  The energies using the ESC and ND potentials
  without and with Coulomb potentials between
  $\alpha$ and $\Xi^-$
   and between triton and $\Xi^-$,
are listed respectively. 
  }
  \label{tab:2}
\begin{center}
  \begin{tabular}{lcccccc}
\hline
\hline
&  &\multispan2 ESC &  
&\multispan2 ND  
\vspace{-3 mm}   
\\
& &\multispan2 \hrulefill  & 
&\multispan2 \hrulefill    \\
&&(without Coulomb) &(with Coulomb) &
&(without Coulomb)  &(with Coulomb) \\
$1^-$ &$B_{\Xi^-}$(MeV)  &$2.24$ &$4.98$ & &$2.23$ &$4.82$\\
&$\Gamma$ $\quad$ (MeV)  &3.95   &4.64 &  &1.38 
&1.66 \\
\hline
$2^-$ &$B_{\Xi^-}$ (MeV) &$3.18$ &$6.08$  & &$1.56$ &$4.06$   \\
&$\Gamma$ $\quad$(MeV)  &4.24   &4.80
 & &0.93   &1.18   \\
\hline
  \end{tabular}
\end{center}
\end{table*}
%=====================================
We listed in Table \ref{tab:2}, the calculated $\Xi^-$-binding energies
($B_{\Xi^-}$) 
of the $1^-_1$ and $2^-_1$ states.
In the case of $G_{ESC}$, the $2^-_1$ state
is obtained at a lower energy than the $1^-_1$ state.
On the other hand, the
use of $G_{ND}$ leads to the opposite order.
In our model, this is because $^{33}S_1$ interaction
for ESC(ND) is more(less) attractive than the
$^{31}S_0$ interaction as shown in Fig.1(c) and (d).
The contribution of the $\Xi^- \alpha$ and $\Xi^- t$
Coulomb forces amounts to about 1.5 MeV.
The conversion widths obtained from
the imaginary part of $G_{ESC}$ is far larger than 
that for $G_{ND}$.
This is because the $^1S_0$ 
$\Lambda \Lambda$-$\Xi N$-$\Sigma \Sigma$ coupling interaction
in ESC is far stronger than that in ND.

We found the appropriate $k_{\rm F}$ parameter
values of 
the effective $\Xi N$ interactions to be
$k_{\rm F}=1.055$ fm$^{-1}$ (ESC)
and $k_{\rm F}=1.025$ fm$^{-1}$(ND),
which 
 are consistent with
the experimental indication
in $^{\:12}_{\Xi^-}$Be.
These interactions provide our basis to investigate
the $A=7$ and $10$ $\Xi^-$ hypernuclei.

%===============================================================
\section{Results for
typical systems composed of $\alpha \Xi^- (^{\: \ 5}_{\Xi^-}$H) and 
$\alpha \alpha \Xi^-(^{\:\ 9}_{\Xi^-}$L\lowercase{i})}
%===============================================================

%%%%%%
Let us study the $\alpha \Xi^-$ and $\alpha \alpha \Xi^-$
systems in order to demonstrate the basic features of the
$\alpha \Xi^-$ interactions.
In the cases of $^{\: \ 7}_{\Xi^-}$H$(\alpha nn\Xi^-$) and
$^{\: 10}_{\Xi^-}$Li$(\alpha \alpha n\Xi^-)$,
the dominant parts of  the $\Xi^-$ binding energies
are given by the $\alpha \Xi^-$ interactions
because of the weak binding of the additional neutrons.
The $\alpha \Xi^-$ interaction is derived by folding
the  $\Xi N$ G-matrix interaction into the wave function of the $\alpha$.
The spin- and isospin-dependent parts, being remarkably
different between ESC and ND, vanish in a folding procedure
involving a spin- and isospin-saturated system such as the $\alpha$. 
Thus, the $\alpha \Xi^-$ interaction is 
determined only by the spin- and isospin-averaged $\Xi N$ 
interaction $\bar G^{(\pm)}(r;k_F)$, where the contribution of
odd-state part $\bar G^{(-)}$ is quite small in the two-body
$\alpha \Xi^-$ system. It should be stressed that $\alpha$-cluster
systems such as $\alpha \Xi^-$ and $\alpha \alpha \Xi^-$ give 
the most basic information on the spin- and isospin-averaged parts 
of $\Xi N$ interactions.
These parts correspond to so-called spin-independent parts
in interactions represented by the $(\sigma \sigma)$,
$(\tau \tau)$ and $(\sigma \sigma)(\tau \tau)$ operators.

Here, it is of vital importance how one chooses the $k_F$ parameters
in our G-matrix interactions.
The parameter $k_F$ specifies the nuclear matter density in 
which the G-matrix interactions are constructed. It is most plausible 
that a corresponding value in a finite system is obtained from an 
average density.
Our basic interactions (ESC and ND) are adjusted so that
the derived G-matrix interactions give rise to reasonable
$\Xi^-$ binding in an $A \sim 12$ system for $k_F=1.0\sim 1.1$
fm$^{-1}$ adequately chosen.
Considering that
the suitable values $k_{\rm F}=1.055$ fm$^{-1}$(ESC)
and $1.025$ fm$^{-1}$(ND) for
$^{\:12}_{\Xi^-}$Be,  it is a modest change to take $k_F=0.9$ fm$^{-1}$ 
in the $A=4 \sim 6$ systems.
In fact, we have had successful prior experience. In 
Ref.\cite{Hiyama97},  we studied the structure of
$^5_{\Lambda}$He, $^9_{\Lambda}$Be and $^{13}_{\Lambda}$C
using the $\Lambda N$ G-matrix interactions, where
 consistent results were obtained by choosing
the $k_{\rm F}$ parameters to be around 0.9 fm $^{-1}$ for 
$^5_{\Lambda}$He and $^9_{\Lambda}$Be and to be around 
1.1 fm$^{-1}$ for $^{13}_{\Lambda}$C.
Here, we take three values of $k_{\rm F}$ parameters for
our $\Xi N$ G-matrix interactions in order to study
$A=7$ and $A=10$ systems:
$k_{\rm F}$=0.9, 1.055 and 1.3 fm$^{-1}$ for ESC,
and $k_{\rm F}$=0.9, 1.025 and 1.3 fm$^{-1}$ for ND.
So, the $k_F$ values for
$A=10$ system are considered to be $k_{\rm F}$ $\sim$
1.0 fm$^{-1}$, while
those for $A=6$ near $k_F=0.9$ fm$^{-1}$.
The unreasonably large value of $k_F=1.3$ fm$^{-1}$, as a trial,
is used only to demonstrate the $k_F$ dependences of the results.

\begin{table}[htbp]
  \caption{The calculated energies of the $1/2^+$ state, 
  $E$, and r.m.s radii, 
  $r_{\alpha {\rm -} \Xi^-}$,
  in the $\alpha \Xi^-(^{\: \ 5}_{\Xi^-}$H)
   system for several values of $k_{\rm F}$.
  The values  in parentheses are energies when the imaginary part of
  the $\alpha \Xi^-$ interactions are switched off.
  The energies are measured from the $\alpha +\Xi^-$ threshold.
  }
  \label{tab:3}
\vspace{0.2cm}
\hspace{1.0cm} (a) $\alpha \Xi^-$(ESC)
\begin{center}
  \begin{tabular}{lcccccc}
\hline
\hline
with   &$k_{\rm F}$(fm$^{-1}$)
&$0.9$ &$1.055$  &$1.3$
 \\
Coulomb &$E$ (MeV) &$-1.36$ &$-0.26$ &$-0.14$ \\
& &$(-1.71)$ &$(-0.57)$ &$(-0.19)$ \\
&$\Gamma$(MeV)  &2.64  &0.86  &0.15  \\
&$r_{\alpha {\rm -} \Xi^-}$(fm) 
&$3.89$ &$6.83$ &$13.6$   \\
\hline
without&$E$ (MeV) &unbound   &unbound &unbound \\
Coulomb & &($-0.64$) & & \\
 &$\Gamma$(MeV)  &-  &-  &-  \\
\hline
  \end{tabular}
\end{center}
\vspace{0.2cm}
\hspace{1.0cm} (b) $\alpha \Xi^-$(ND)
\begin{center}
  \begin{tabular}{lcccccc}
\hline
\hline
with  &$k_{\rm F}$(fm$^{-1}$)
 &$0.9$ &$1.025$  &$1.3$
 \\
Coulomb &$E$ (MeV) &$-0.57$ &$-0.32$ &$-0.15$ \\
 & &$(-0.57)$ &$(-0.32)$ &$(-0.16)$ \\
&$\Gamma$(MeV)  &0.16   &0.06  &0.004  \\
&$r_{\alpha {\rm -} \Xi^-}$(fm) 
&$6.87$ &$9.82$ &$15.65$   \\
\hline
without &$E$ (MeV) &unbound &unbound &unbound \\
Coulomb &$\Gamma$(MeV)  &-  &-  &-  \\
\hline
  \end{tabular}
\end{center}
\end{table}

In Table \ref{tab:3}, we show the calculated  
energies and r.m.s radii for the $\alpha$$\Xi^-$ system 
for three  $k_F$ values 
of the $\Xi N$ G-matrix interactions.
Of course, Coulomb bound ($\Xi^-$-atomic) states are obtained, 
even if the strong interactions are switched off. 
If a bound state is obtained without the Coulomb interaction, 
this state is called a nuclear-bound state.
When a nuclear-unbound state becomes bound with help of
the the attractive Coulomb interaction, 
such a state is called a Coulomb-assisted bound state.
Table \ref{tab:3} summarizes our results that
in each case the lowest state is
found to be a 
Coulomb-assisted bound state, namely
there appears no nuclear-bound state.
It is noted, in this Table, that a nuclear-bound state is
obtained in the case of $G_{ESC}(k_F=0.9)$ if its imaginary
part is switched off:
Though the real part of the $\alpha$$\Xi^-$ interaction
for $G_{ESC}(k_F=0.9)$ is attractive enough to give a 
nuclear-bound state, the strong imaginary part makes
the resulting state nuclear-unbound.
In any case, the spin- and isospin-averaged even-state part 
$\bar G^{(+)}$ for $G_{ESC}$ is far more attractive than 
the corresponding part of $G_{ND}$.

In  Table \ref{tab:5}, we list the calculated
results 
for the $\alpha \alpha \Xi^-$ system.
It should be noted, here, that nuclear-bound states are
obtained in both cases of $G_{ESC}$ and $G_{ND}$
unless an unreasonably large value of $k_F$ is chosen.
The calculated  energies for $G_{ESC}$ are naturally 
larger than 
those for $G_{ND}$.
In the $\alpha \alpha \Xi^-$ system, however,
$\bar G^{(-)}$ contributes significantly. 
It is remarked that the 
odd-state interaction in $G_{ND}$
is far more attractive than that in $G_{ESC}$, which works
to reduce the difference between both potentials in 
the $\alpha \alpha \Xi^-$ system.

\begin{table}[htbp]
  \caption{The calculated energies of the $1/2+$ state,
  $E$, in the $\alpha \alpha \Xi^-(^{\:\ 9}_{\Xi^-}$Li) system for
several values of $k_{\rm F}$.
The values in parentheses are energies when the
imaginary part of the $\alpha \Xi^-$ interaction are switched off.
The energies are measured from the
$\alpha \alpha \Xi^-$ three-body breakup threshold.}
  \label{tab:5}
\vspace{0.2cm}
\hspace{1.0cm} (a) $\alpha \alpha \Xi^-$ (ESC)
\begin{center}
\begin{tabular}{lcccccc}
\hline
\hline
with  &$k_{\rm F}$(fm$^{-1}$)
 &$0.9$ &$1.055$  &$1.3$
 \\
Coulomb &$E$ (MeV) &$-4.81$  &$-2.23$ &$-0.83$ \\
& &$(-5.17)$  &$(-2.57)$ &$(-1.04)$ \\
&$\Gamma$(MeV)  &$5.01$   &$2.89$  &1.18  \\
\hline
without &$E$ (MeV) &$-2.54$  &$-0.41$ &unbound \\
Coulomb & &$(-2.94)$ &$(-0.77)$  & \\
&$\Gamma$(MeV)  &$4.48$   &$2.18$  &-  \\
\hline
  \end{tabular}
\end{center}
\vspace{0.2cm}
\hspace{1.0cm} (b) $\alpha \alpha \Xi^-$ (ND)
\begin{center}
  \begin{tabular}{cccccc}
\hline
\hline
with  &$k_{\rm F}$(fm$^{-1}$)
  &$0.9$ &$1.025$  &$1.3$
 \\
Coulomb &$E$ (MeV) &$-2.87$  &$-1.82$ &$-0.79$ \\
& &$(-2.89)$  &$(-1.83)$  &$(-0.79)$ \\
&$\Gamma$(MeV)  &$0.58$   &$0.3$  &0.06  \\
\hline
without &$E$ (MeV) &$-1.02$ &$-0.25$ &unbound \\
Coulomb& &$(-1.03)$  &$(-0.26)$ &  \\
&$\Gamma$(MeV)  &$0.45$   &0.20  &-  \\
\hline
  \end{tabular}
\end{center}
\end{table}

One notices in the Tables,
that the decay widths for $G_{ESC}$ are much 
larger than those for $G_{ND}$, when they are calculated for
the same value of $k_F$.
This is because the imaginary part of $G_{ESC}$ is stronger than
that of $G_{ND}$. The difference of the imaginary parts 
originates mainly from the different strengths of $^{11}S_0$
$\Lambda \Lambda$-$\Xi N-\Sigma \Sigma$ coupling interactions in ESC and ND.

As mentioned before, 
noting that the choice of the
$k_F$ value
$\sim 0.9$ fm$^{-1}$($\alpha \Xi^-$)
and  $\sim 1.0$ fm$^{-1}$ $(\alpha \alpha \Xi^-)$,
are reasonable, respectively,
we can expect the existence
of nuclear-bound states,
especially, in the latter case.
Thus, we can say that observations of $\alpha \Xi^-$ and
$\alpha \alpha \Xi^-$ systems certainly provide
information about
spin-independent parts of the $\Xi N$ interactions.
In reality,
 however, there are no corresponding nuclear targets  
to produce the above systems by the $(K^-,K^+)$ reaction.
As their actual substitutes, in the
following, we investigate the structures
of $^{\:\ 7}_{\Xi^-}$H($\alpha  nn\Xi^-)$ and
$^{\:10}_{\Xi^-}$Li($\alpha \alpha n \Xi^-)$
having additional neutron(s),
and propose  to perform 
the $^7$Li$(K^-,K^+)$ and
$^{10}$B$(K^-,K^+)$ reaction 
experiments with available targets.

%===============================================================
\section{$A=7$ and $A=10$ $\Xi^-$ hypernuclei}
%===============================================================

Here, we study $^{\:\ 7}_{\Xi^-}$H and $^{\: 10}_{\Xi^-}$Li on the basis 
of $\alpha  nn \Xi^-$  and $\alpha \alpha  n \Xi^-$ four-body
cluster models, respectively.
In cluster-model studies, it is essential that
interactions among cluster subunits be given 
consistently with respect to the corresponding threshold energies.
Namely, low-energy bound-state energies and 
scattering phase shifts of $\alpha n$, $\alpha \alpha$, 
,$\alpha nn$, and $\alpha \alpha n$ subsystems should be reproduced 
reasonably by the corresponding interactions.
We emphasize that these severe constraints
are correctly satisfied in the present models, 
as mentioned below.

%===============================================================
\subsection{Model and Interactions}

\begin{figure}[htb]
\begin{center}
\epsfig{file=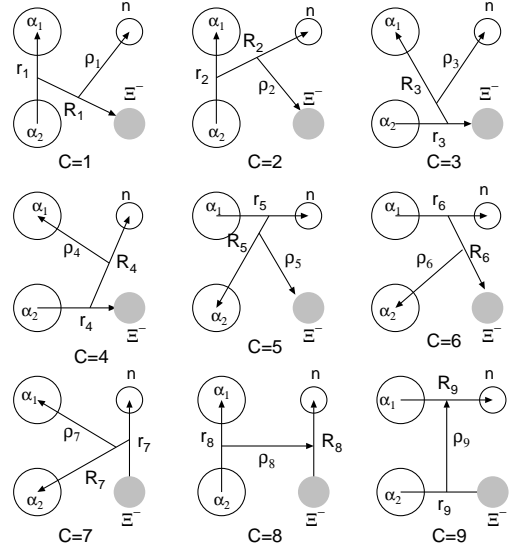,scale=0.35}
\end{center}
\caption{Jacobian coordinates for all the rearrangement channels
($c=1 \sim 9$) of the $\alpha \alpha n\Xi^-(^{\: 10}_{\Xi^-}$Li)  
four-body system. Two $\alpha$ clusters are to be
symmetrized. In the case of the $\alpha nn \Xi^-(^{\:\ 7}_{\Xi^-}$H)  four-body
system, the two $\alpha$ clusters are replaced by
two neutrons, and the neutron is replaced by
an $\alpha$ cluster.}
\label{fig:b10x}
\end{figure} 

\begin{figure*}[htb]
\begin{minipage}[t]{8 cm}
%\begin{minipage}{0.6\linewidth}
%\scalebox{0.55}
\epsfig{file=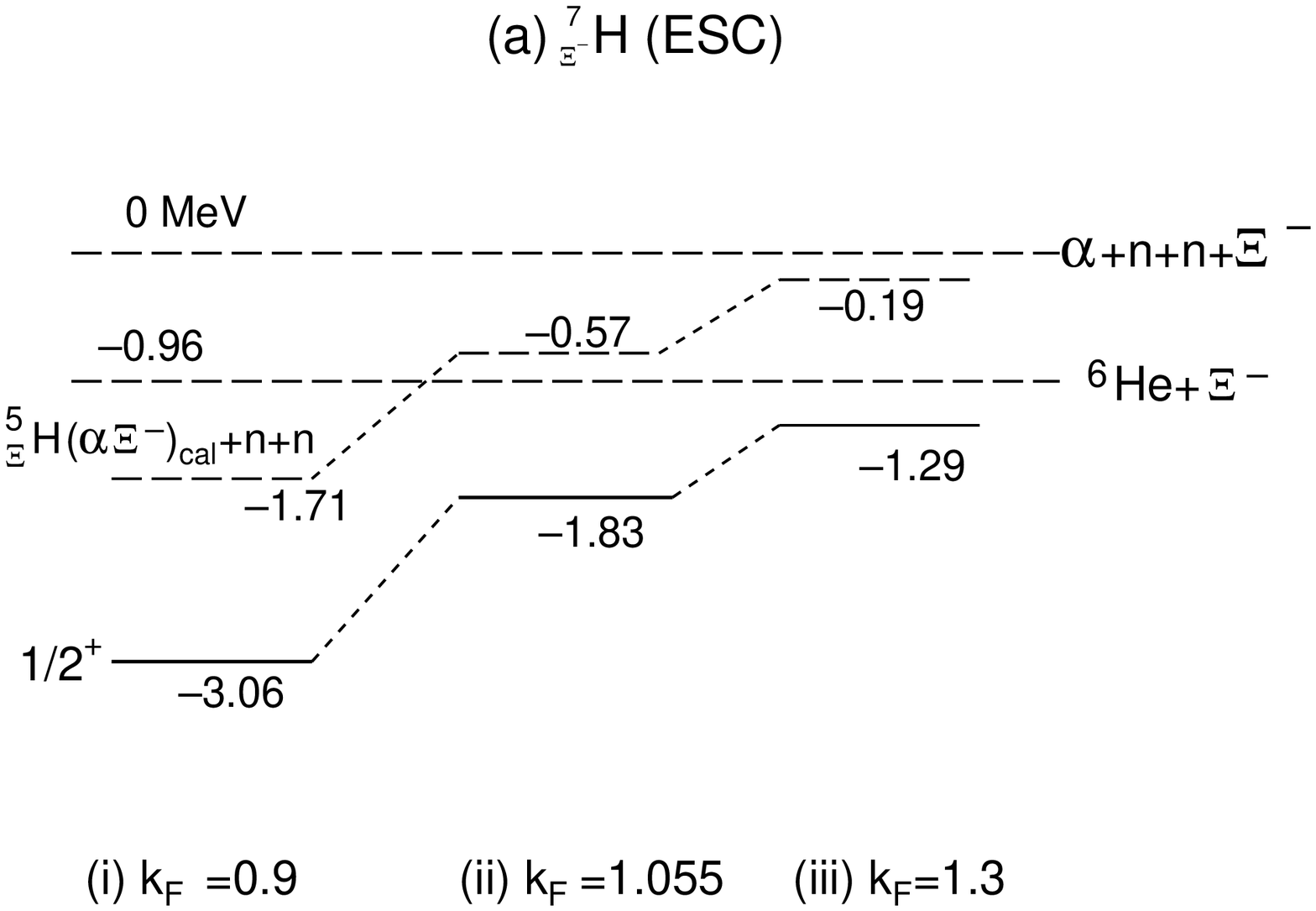,scale=0.41}
%{\includegraphics{li7xevel.eps}}
\end{minipage}
\hspace{\fill}
\begin{minipage}[t]{8.5 cm}
%\begin{minipage}{0.6\linewidth}
%\scalebox{0.55}
%{\includegraphics{li7xevel-nd.eps}}
\epsfig{file=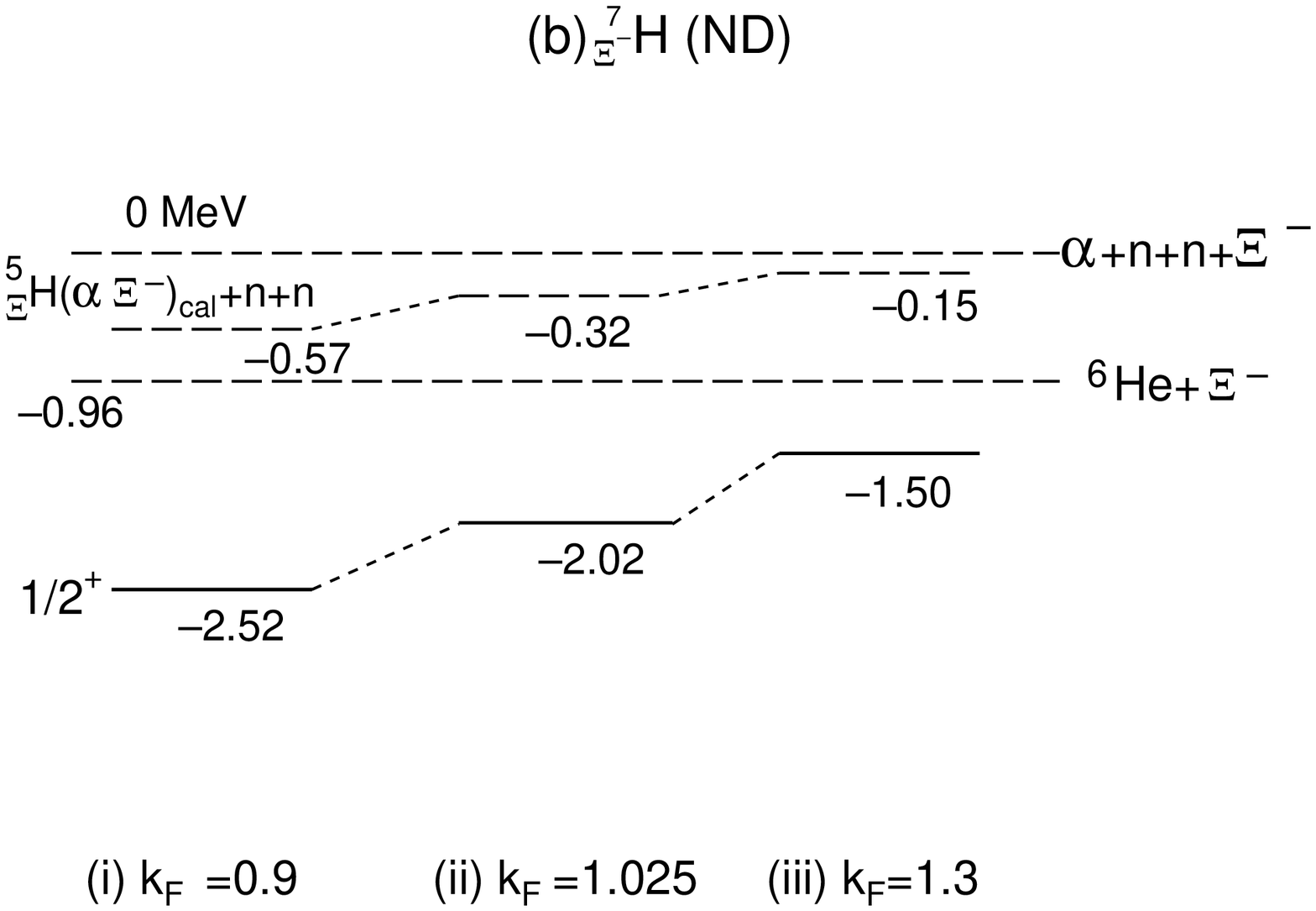,scale=0.41}
\end{minipage}
\caption{(a)Calculated energy levels of
$^{\: \ 7}_{\Xi^-}$H for three  $k_{\rm F}$ values
 using ESC.
(b) Calculated energy levels of
$^{\: \ 7}_{\Xi^-}$H for for three  $k_{\rm F}$ values
using ND.
The energies are shown when the imaginary part of the
$\alpha \Xi^-$ interaction is switched off.
The energies are measured from the $\alpha+n+n+\Xi^-$ breakup threshold.
The dashed lines are threshold.
}
\label{fig:li7xlevel}
\end{figure*}

For $^{\:\ 7}_{\Xi^-}$H and $^{\:10}_{\Xi^-}$Li,
all nine sets of the Jacobian coordinate of
the four-body systems are shown in Fig.~\ref{fig:b10x}, respectively.
The total Hamiltonian and the Schr\"{o}dinger equation
are given by 
\begin{eqnarray}
 && ( H - E ) \, \Psi_{JM}(^{\:\ 7}_{\Xi^-}{\rm H},\:\:
 ^{\: 10} _{\Xi^-}{\rm Li})  = 0 \ , \\
 &&  H 
 =T+\sum_{a,b}V_{a b}
      +V_{\rm Pauli} \ ,
\label{eq:hamiltonian}
\end{eqnarray}
where $T$ is the kinetic-energy operator,
$V_{ab}$ is the interaction between the 
constituent particle $a$ and $b$, and
the $V_{\rm Pauli}$ is 
the Pauli projection
operator given by Eq(\ref{eq:ocm}).
The total wavefunction
is described as a sum of amplitudes
of the rearrangement channels $(c=1 \sim 9)$
of Fig.~\ref{fig:b10x} in the $LS$ coupling
scheme:
\begin{eqnarray}
      \Psi_{JM\:,TT_z}\!\!&(&\!\!^{\:\ 7}_{\Xi^-}{\rm H})
       =  \sum_{c=1}^{9}
      \sum_{n,N,\nu}  \sum_{l,L,\lambda}
       \sum_{S,\Sigma,I,K}
       C^{(c)}_{nlNL\nu\lambda S\Sigma IK} \nonumber  \\
      &  \times & {\cal A}_N
      \Big[
       \Phi(\alpha) \big[\chi_{\frac{1}{2}}(\Xi^-)
     \big[ \chi_{\frac{1}{2}}(n_1)
       \chi_{\frac{1}{2}}(n_2)
       \big]_S \big]_\Sigma  \nonumber \\
  & \times  &    \big[ \big[ \phi^{(c)}_{nl}({\bf r}_c)
         \psi^{(c)}_{NL}({\bf R}_c)\big]_I
        \xi^{(c)}_{\nu\lambda} (\mbox{\boldmath $\rho$}_c)
        \big]_{K}  \Big]_{JM}\: \nonumber \\
        & \times  &
        \big[\eta_{\frac{1}{2}}(\Xi^-)
        \big[ \eta_{\frac{1}{2}}(n_1)
       \eta_{\frac{1}{2}}(n_2)
       \big]_t \big]_{T,T_z}    \;  ,
\label{eq:a7xiwavefunction}
\end{eqnarray}
\begin{eqnarray}
      \Psi_{JM,\:TT_z}\!\!&(&\!\!^{\: 10}_{\Xi^-}{\rm Li})
       =  \sum_{c=1}^{9}
      \sum_{n,N,\nu}  \sum_{l,L,\lambda}
       \sum_{S,I,K}
       C^{(c)}_{nlNL\nu\lambda S IK} \nonumber  \\
      &  \times & {\cal S}_\alpha
      \Big[
       \Phi(\alpha_1)\Phi(\alpha_2) 
     \big[ \chi_{\frac{1}{2}}(n)
       \chi_{\frac{1}{2}}(\Xi^-)
       \big]_S   \nonumber \\
  & \times  &    \big[ \big[ \phi^{(c)}_{nl}({\bf r}_c)
         \psi^{(c)}_{NL}({\bf R}_c)\big]_I
        \xi^{(c)}_{\nu\lambda} (\mbox{\boldmath $\rho$}_c)
        \big]_{K}  \Big]_{JM}  \nonumber \\
   & \times  &    \big[\eta_{\frac{1}{2}}(n) 
        \eta_{\frac{1}{2}}(\Xi^-) \big]_{T,T_z} \;  .
%(2.3)
\label{eq:wavefunction}
\end{eqnarray}
Here the operator $\cal{A}_N$ stands for
antisymmetrization between the two neutrons.
${\cal S}_\alpha$, $\chi_{\frac{1}{2}}(\Xi^-)$
and $\eta_{\frac{1}{2}}(\Xi^-)$ are
defined already in  Sec.III.

The Pauli principle involving nucleons belonging to
$\alpha$ and $x(=n,\alpha)$ is
taken into account by the orthogonality condition
model (OCM) \cite{Saito69}.
 The forbidden states in
 Eq.(\ref{eq:ocm}) are $f=0S$ for $n$
and $f={0S,1S,0D}$ for $x=\alpha$.

We employ the $V_{\alpha N}$ potential given in Ref.\cite{Kanada79}
and the AV8 potential \cite{Pudliner97} for the two-neutron parts.
The $\alpha nn$ ($\alpha \alpha n)$ binding energy derived from these potentials
is less(over) bound by about 0.3 MeV (1 MeV) 
in comparison with the observed value.
Then, in calculations of the $\alpha  nn \Xi^-$ 
and $\alpha \alpha n \Xi^-$ four-body model,
the central part of $V_{\alpha n}$ is adjusted
so as to reproduce the observed ground state of $^6$He
and $^9$Be.
The $V_{\alpha \alpha}$ and $V_{\alpha \Xi^-}$ are the same 
as those in $\alpha \alpha t \Xi^-$ four-body calculations.
As for the $\Xi^- n$ parts,
we employ the simple three-range Gaussian potentials
derived from ESC and ND.
The details of these potentials were already
mentioned in Sec.II.  Thus, in our treatments of
$\alpha  nn\Xi^-$ and $\alpha \alpha  n \Xi^-$ four-body systems,
ground-state energies of all subsystems of $\alpha nn$ and $\alpha \alpha n$
are reproduced well.

%===============================================================
\subsection{Results for $^{\: \ 7}_{\Xi}$H ($\alpha  nn \Xi^-$)}

\begin{figure*}[htb]
\begin{minipage}{0.35\linewidth}
\scalebox{0.35}
{\includegraphics{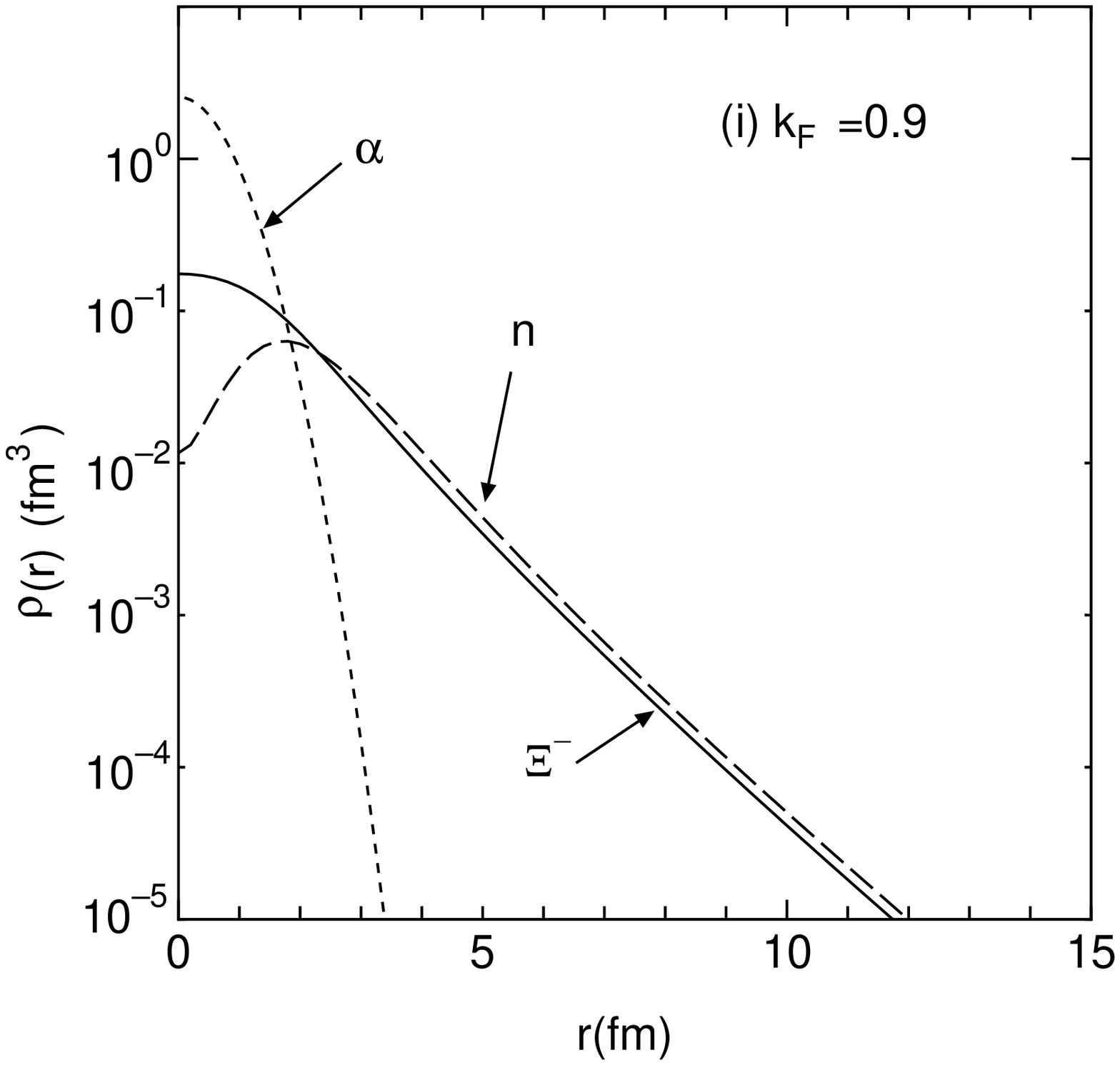}}
\end{minipage}
\begin{minipage}{0.31\linewidth}
(a)$^7_{\Xi^-}$H(ESC)
\vspace{0.5cm}
\scalebox{0.35}
{\includegraphics{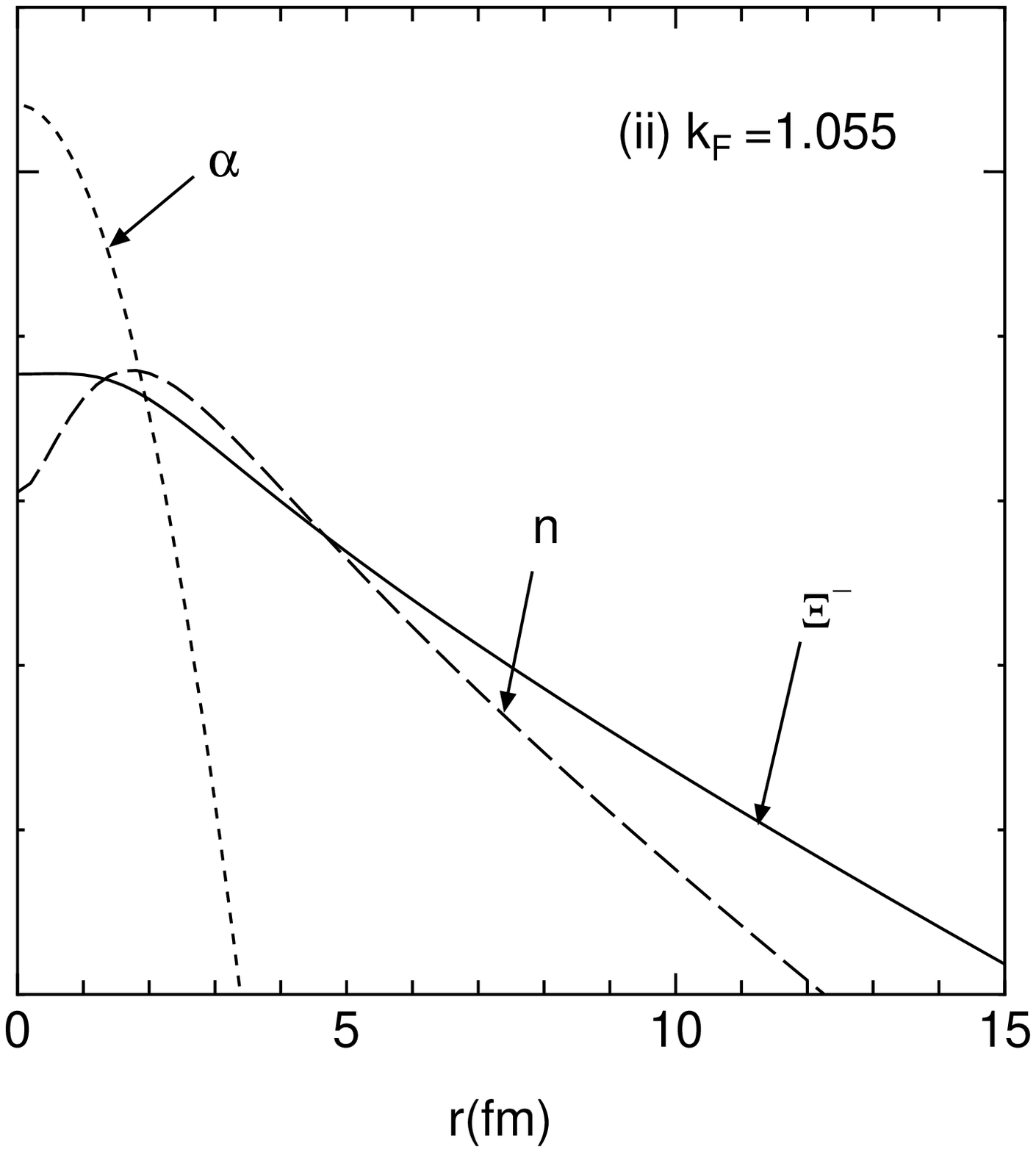}}
\end{minipage}
\begin{minipage}{0.31\linewidth}
\scalebox{0.35}
{\includegraphics{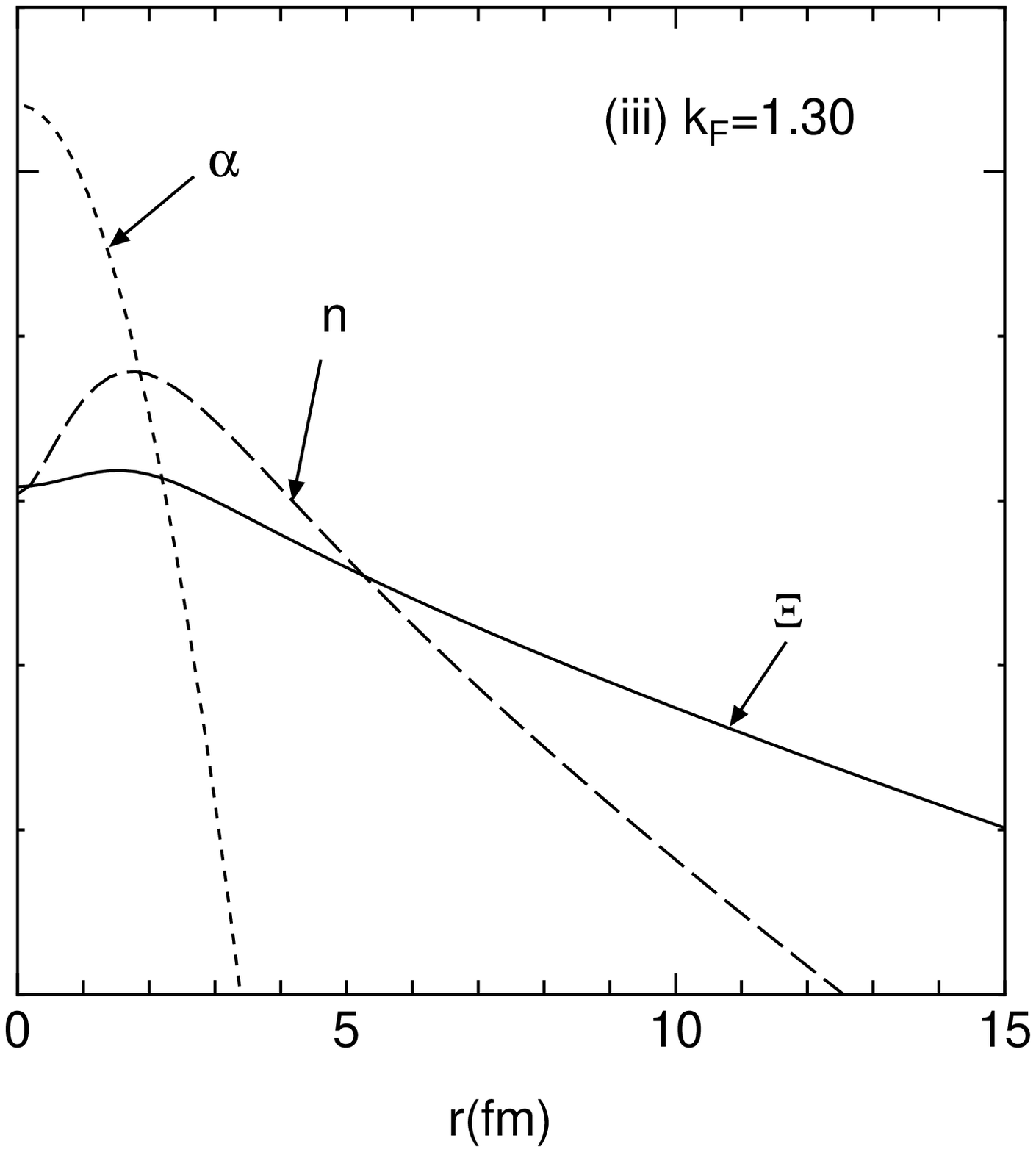}}
\end{minipage}
\begin{minipage}{0.35\linewidth}
\scalebox{0.35}
{\includegraphics{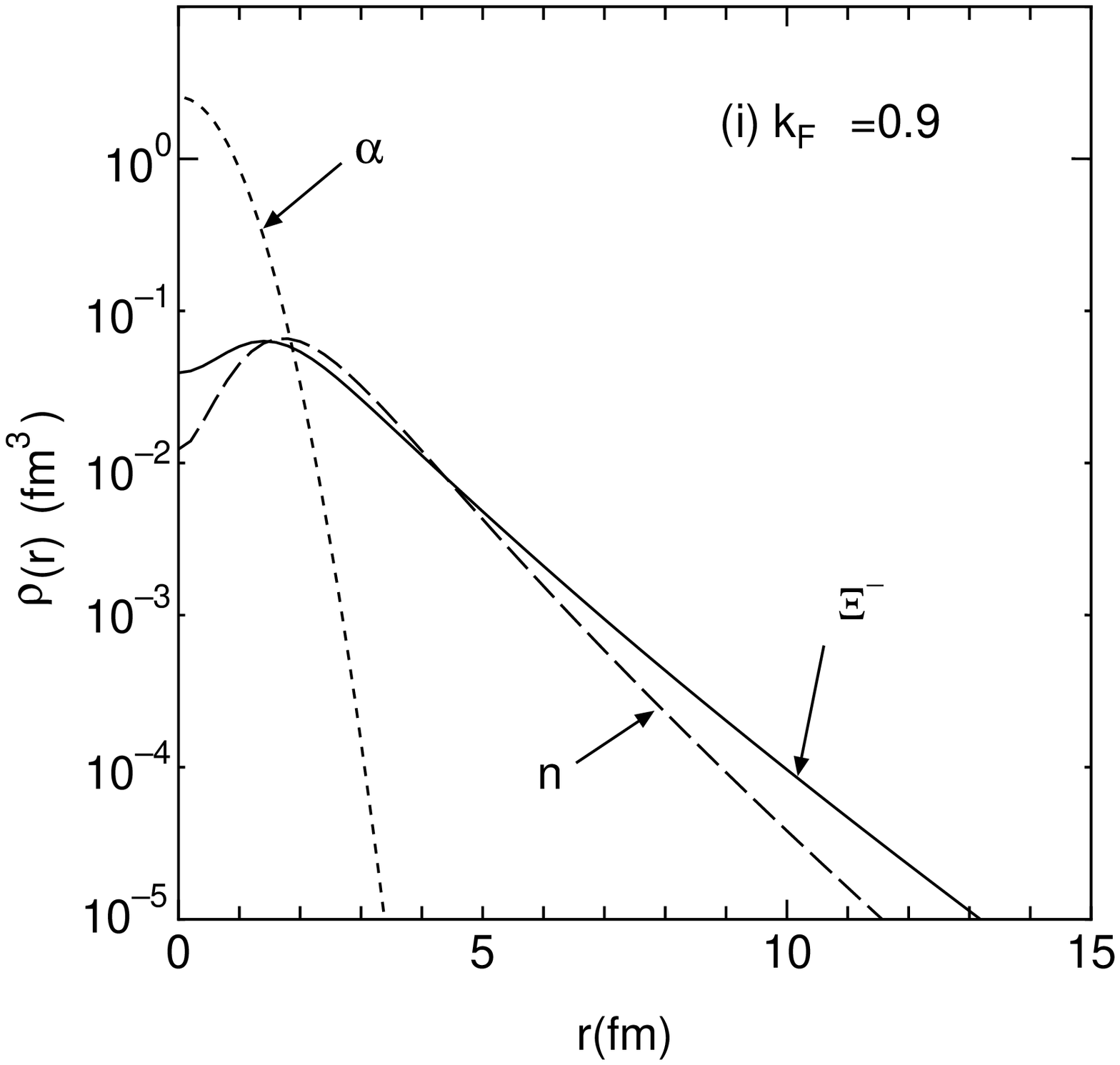}}
\end{minipage}
\begin{minipage}{0.31\linewidth}
(b)$^7_{\Xi^-}$H(ND)
\vspace{0.5cm}
\scalebox{0.35}
{\includegraphics{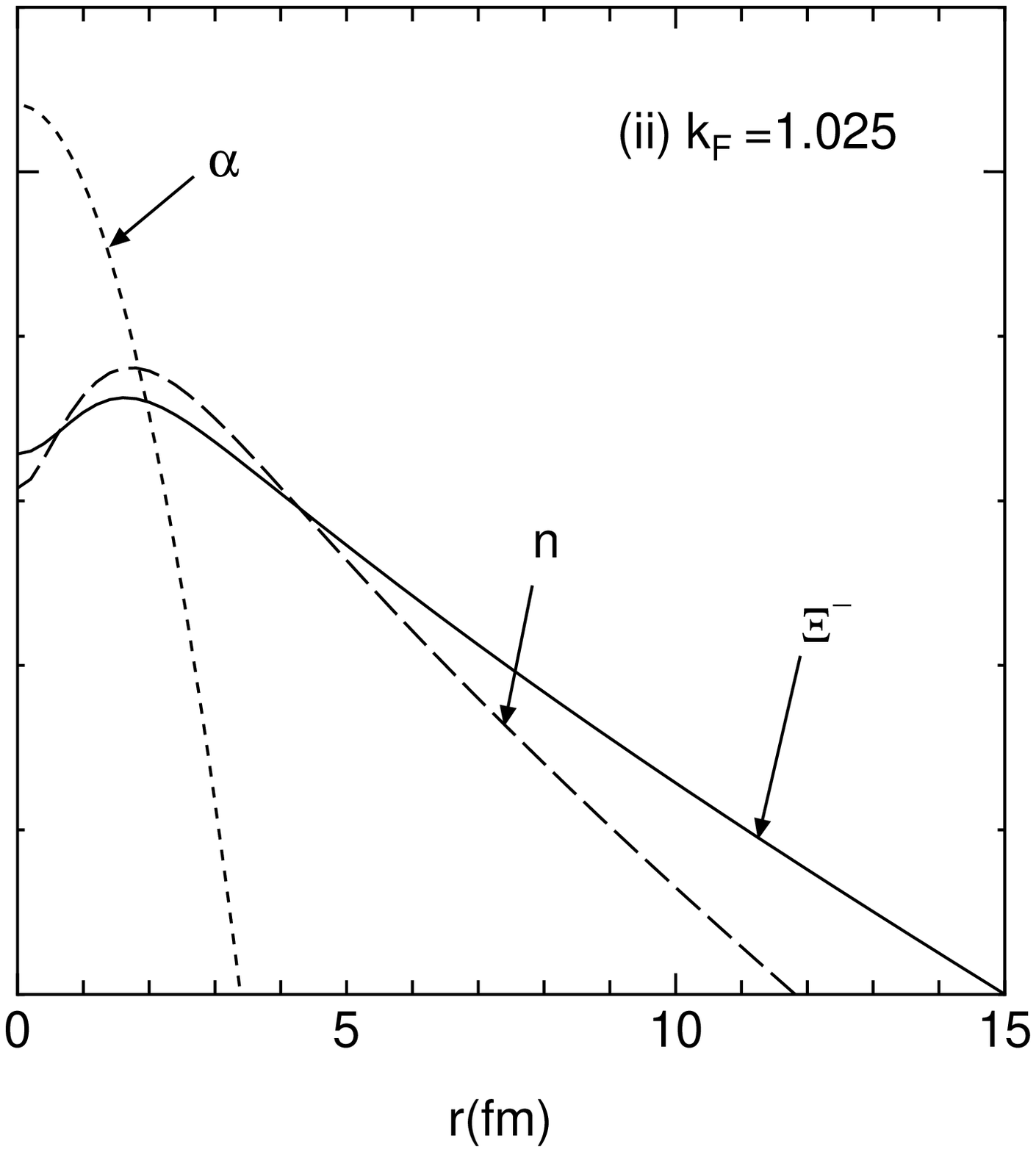}}
\end{minipage}
\begin{minipage}{0.31\linewidth}
\scalebox{0.35}
{\includegraphics{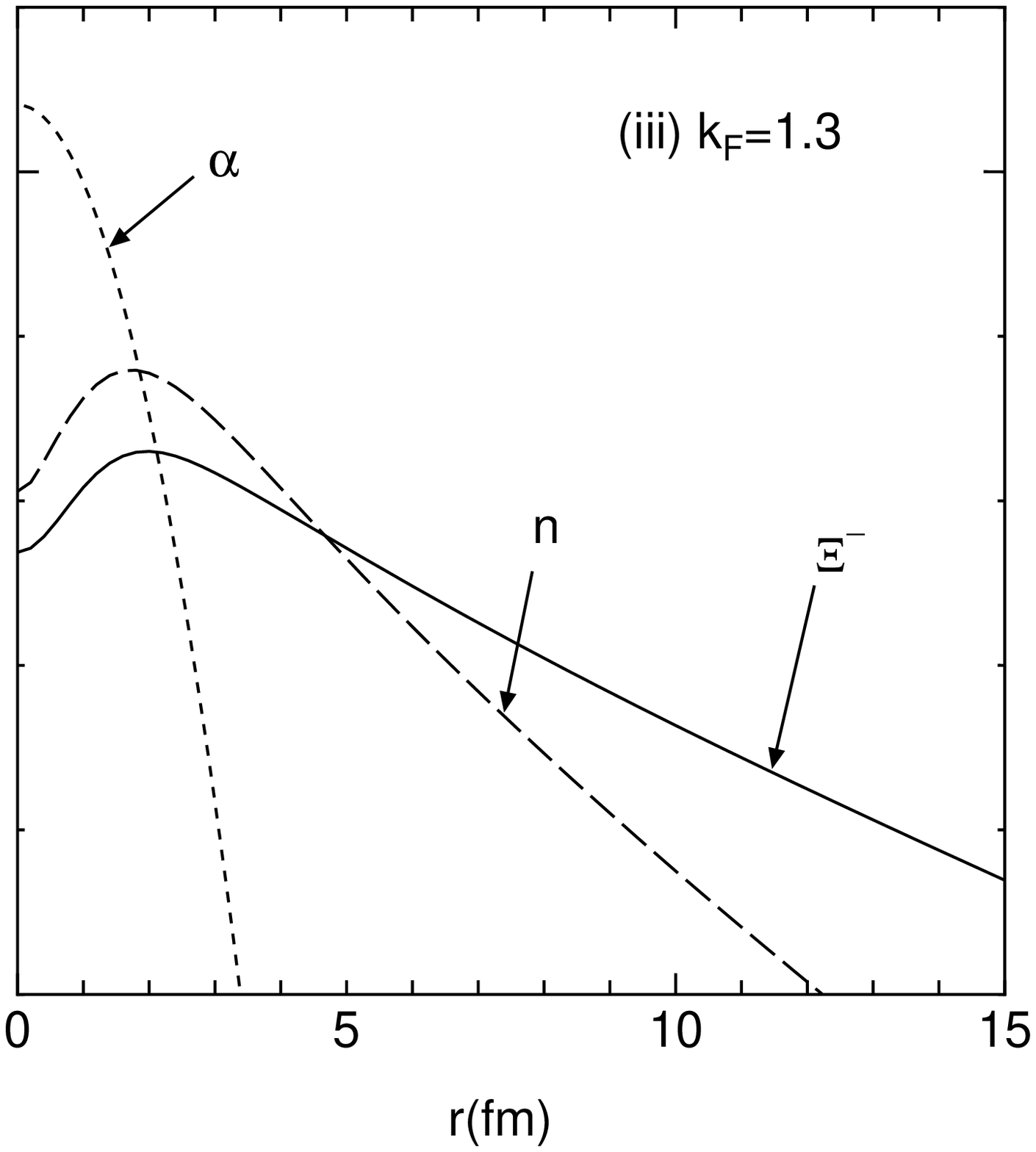}}
\end{minipage}
\caption{(a)Calculated density distribution of
$\alpha$, $\Xi^-$ and a valence neutron
for three  $k_{\rm F}$ values using ESC.
(b)Calculated density distribution of
$\alpha$, $\Xi^-$ and a valence neutron
for three  $k_{\rm F}$ values using ND.
The wavefunctions of $\alpha \Xi^-$ without the
imaginary part of the $\alpha \Xi^-$ interaction are used.}
\label{fig:li7den}
\end{figure*}

%%%%%%%%%%%%

Here we describe the results of the
four-body calculations for
$^{\:\ 7}_{\Xi^-}$H$(\alpha nn \Xi^-)$ with
$(T,J^\pi)=(3/2,1/2^+)$. 
The basic question is whether this state is bound or not:
The $^6$He core is composed of an $\alpha$ and 
two weakly-bound ('halo') neutrons. Due to the weakness of the
$\Xi^- n$ interaction,
the binding between $^6$He and $\Xi^-$ is 
to a large extent  determined by the $\alpha$$\Xi^-$ interaction.

The calculated energies 
in the $1/2^+$ ground state are
demonstrated in Fig.~\ref{fig:li7xlevel}
as a function of $k_{\rm F}$, 
for the two $\Xi N$ potential models without
the imaginary part of the $\alpha \Xi^-$ interaction.
These $1/2^+$ states are composed of the ground-state $0^+$
configuration of $^6$He coupled with the $0s$-state $\Xi^-$ particle.
The Coulomb interactions between $\alpha$ and $\Xi^-$
are taken into account.
In the figure, the dashed lines show the positions of 
threshold energies of $\alpha+n+n+\Xi^-$, $^6$He$+\Xi^-$
and $^{\:\ 5}_{\Xi^-}$H$(\alpha \Xi^-)_{\rm cal}+n+n$,
respectively. One should 
be aware that the $^{\:\ 5}_{\Xi^-}$H$(\alpha \Xi^-)_{\rm cal}+n+n$
threshold energy depends on the $k_{\rm F}$ value
of the adopted 
$\Xi N$ interactions.
This situation is unavoidable, because the calculated energies
for $^{\:\ 5}_{\Xi^-}$H have to be used instead of the unknown 
experimental value.
We see that in the case (i) $k_{\rm F}=0.9$ fm$^{-1}$
with ESC
the lowest threshold is $^{\:\ 5}_{\Xi^-}$H($\alpha \Xi^-)_{\rm cal}+n+n$,
and in the other cases the $^6{\rm He}+\Xi^-$ 
threshold is lower than the $^{\:\ 5}_{\Xi^-}$H($\alpha \Xi^-)_{\rm cal}+n+n$ 
threshold. On the other hand, in all $k_{\rm F}$
 cases 
 with ND the lowest threshold is $^6{\rm He}+\Xi^-$.
The order of the
$^{\:\ 5}_{\Xi^-}$H($\alpha \Xi^-)_{\rm cal}+n+n$ and
$^6{\rm He}+\Xi^-$ threshold is determined by the
competition between $\alpha$-$\Xi^-$ correlation
and the $\alpha$ -$(nn)$ correlation.

More detailed results are given in Table \ref{tab:4},
where  the calculated values of the conversion
widths $\Gamma$ and the $\alpha$$\Xi^-$ and
$\alpha n$ r.m.s. radii are also listed. 

Now, let us compare the results for ESC and ND
in the cases in which the imaginary part of the
$\alpha \Xi^-$ interaction is switched off.
As found in Table \ref{tab:3} (values in parentheses),
the obtained $\alpha \Xi^-$ states for ESC are more bound 
than those for ND ($-1.71$ MeV vs. $-0.57$ MeV for
$k_{\rm F}=0.9$ fm$^{-1}$).
In the $\alpha nn \Xi^-$ system, however, the energy difference
between ESC and ND becomes small
in comparison with that in the $\alpha \Xi^-$ system
($-3.06$ MeV vs. $-2.52$ MeV for $k_{\rm F}=0.9$ fm$^{-1}$),
as shown in Fig.~\ref{fig:li7xlevel} and Table \ref{tab:4}
(values in parentheses).
This is because the $^{31}S_0$ and $^{33}S_1$ 
$n\Xi^-$ interactions of ND are more attractive
than those of ESC, 
as shown in Fig.~\ref{Volint}.
The stronger $n\Xi^-$ attraction in ND has
the effect of a larger 
reduction of the value of ${\bar r}_{\alpha {\rm -} \Xi^-}$
when one goes
from the $\alpha \Xi^-$ system to the $\alpha nn \Xi^-$ system.

Let us discuss the structure of $^{\:\ 7}_{\Xi^-}$H.
In the case of $k_{\rm F}=0.9$ fm$^{-1}$ for ESC, 
the lowest threshold is
$^{\:\ 5}_{\Xi^-}$H($\alpha \Xi^-)_{\rm cal}+n+n$.
Then, the $\Xi^-$ particle is bound to the $\alpha$ particle mostly 
in the $0s$ orbit, and the two valence  neutrons are coupled to the
$\alpha \Xi^-$ subsystem.
In fact, as shown in Table \ref{tab:4},
${\bar r}_{\alpha {\rm -} \Xi^-}$ is shorter than  
${\bar r}_{\alpha {\rm -} n}$ in this case.
In other cases, the
two valence  neutrons are bound to the $\alpha$ core, and the
$\Xi^-$ particle is coupled to the $\alpha nn(^6$He) subsystem,
corresponding to where the ${\bar r}_{\alpha {\rm -} \Xi^-}$ values 
are larger than the ${\bar r}_{\alpha {\rm -} n}$ values.

In order to see the structure of the $^{\:\ 7}_{\Xi^-}$H 
system visually, we draw the density distributions of 
 $\Xi^-$ (solid curves) and valence neutrons (dashed curves)
in Fig.~\ref{fig:li7den}(a) and (b)
for ESC and ND, respectively.
For comparison, also a single-nucleon density in the 
$\alpha$ core is shown by a dotted curve in each case.
It turns out, here, that as the binding energies of
$^{\:\ 7}_{\Xi^-}$H become smaller,
the $\Xi^-$ density distribution has a longer tail.
As is well known, $^6$He is a neutron-halo nucleus.
It is interesting here to see the overlapping of the $\Xi^-$ 
distribution with the halo-neutron distribution.
In the case of $k_{\rm F}=0.9$ fm$^{-1}$
for ESC, since the lowest threshold is 
$^{\:\ 5}_{\Xi^-}{\rm H}(\alpha \Xi^-)_{\rm cal}+n+n$,
the density of the $\Xi^-$ particle has a shorter-ranged tail 
than that of the two valence neutrons,
but is extended significantly away form the $\alpha$ core.
This situation can be visualized as 
three layers of matter distribution,
the $\alpha$ core, a $\Xi^-$ skin, and neutron halo.
When the lowest breakup threshold is $^6{\rm He}+\Xi^-$,
the $\Xi^-$ density is  longer-ranged than that of
the valence neutrons due to the weaker binding of the $\Xi^-$ particle.
Then, the density distribution of $^{\:\ 7}_{\Xi^-}$H shows the  
three layers of the $\alpha$ core, neutron halo, and $\Xi^-$ halo.
Namely, a double-halo structure of neutrons and $\Xi^-$ exists, 
in which the attractive Coulomb interaction plays an essential role.
These features can be considered as
new forms in baryon many-body systems.

\begin{table}[htbp]
  \caption{The calculated binding energies of $1/2^+$, $E$ and r.m.s radii, 
  ${\bar r}_{\alpha {\rm -} \Xi^-}$ and ${\bar r}_{\alpha {\rm -} n}$ ,
  in the $^{\:\ 7}_{\Xi^-}{\rm H}(\alpha  nn\Xi^-$)
   system for several values of $k_{\rm F}$.
  The values  in parentheses are energies when the imaginary part of the
  $\alpha \Xi^-$ interactions are switched off.
  The energies are measured from the $\alpha +\alpha+n+ \Xi^-$ threshold.
  }
  \label{tab:4}
\vspace{0.2cm}
\hspace{1.0cm} (a) $^{\:\ 7}_{\Xi^-}$H(ESC)
\begin{center}
  \begin{tabular}{lcccccc}
\hline
\hline
with  &$k_{\rm F}$(fm$^{-1}$)
 &$0.9$ &$1.055$  &$1.3$
 \\
Coulomb &$E$ (MeV) &$-2.76$ &$-1.63$ &$-1.22$ \\
& &$(-3.06)$  &$(-1.83)$ &$(-1.29)$  \\
&$\Gamma$(MeV)  &2.64   &1.15  &0.31  \\
&${\bar r}_{\alpha {\rm -} \Xi^-}$(fm) 
&$3.68$ &$5.58$ &$9.92$   \\
&${\bar r}_{\alpha {\rm -} n}$(fm) 
&$4.04$ &$4.11$ &$4.19$   \\
\hline
without &$E$ (MeV) &$-1.68$ &unbound  &unbound \\
Coulomb& &$(-1.96)$ &$(-1.09$) &(unbound) \\
&$\Gamma$(MeV)  &2.09   &-  &- \\
\hline
  \end{tabular}
\end{center}
\vspace{0.2cm}
\hspace{1.0cm} (b) $^{\:\ 7}_{\Xi^-}$H(ND)
\begin{center}
  \begin{tabular}{lcccccc}
\hline
\hline
with  &$k_{\rm F}$(fm$^{-1}$)
 &$0.9$ &$1.025$  &$1.3$
 \\
Coulomb &$E$ (MeV)  &$-2.51$ &$-2.01$ &$-1.50$ \\
& &$(-2.52)$  &$(-2.02)$  &$(-1.50)$ \\
&$\Gamma$(MeV)  &0.27   &0.15  &0.032  \\
&${\bar r}_{\alpha {\rm -} \Xi^-}$(fm) 
&$4.48$ &$5.35$ &$7.55$   \\
&${\bar r}_{\alpha {\rm -} n}$(fm) 
&$3.92$ &$3.99$ &$4.11$   \\
\hline
without &$E$ (MeV) &$-1.62$  &$-1.26$ &unbound \\
Coulomb& &$(-1.63)$  &$(-1.26$) &(unbound) \\
&$\Gamma$(MeV)  &$0.22$   &$0.10$  &-  \\
\hline
  \end{tabular}
\end{center}
\end{table}

 Table \ref{tab:4}
lists the binding energies of the $^{\:\ 7}_{\Xi^-}$H system
calculated with and without the Coulomb interaction for
each $k_{\rm F}$ value.
For ESC ($k_{\rm F}=0.9$ fm$^{-1}$) and 
ND ($k_{\rm F}=0.9$ and 1.025 fm$^{-1}$)
the ground states of $^{\:\ 7}_{\Xi^-}$H are found to be weakly
bound states, when the Coulomb interactions are switched off.
%On the other hand, for ESC ($k_{\rm F}=0.9$ fm$^{-1}$), 
%since the $^{\:\ 7}_{\Xi^-}$Li is bound system without
%Coulomb interaction.
Therefore, the $^{\:\ 7}_{\Xi^-}$H systems 
are seen to have
nuclear-bound states, if we take 
reasonable values $k_{\rm F} < 1$ fm$^{-1}$.
This means that 
an experimental finding of a $^{\:\ 7}_{\Xi^-}$H bound state 
indicates the existence of an $\alpha \Xi^-$  bound state
in which the even-state spin-independent part of the 
$\Xi N$ interaction is substantially attractive.
This statement is almost independent of the interaction model.

\begin{figure*}[htb]
\begin{minipage}[t]{8 cm}
\epsfig{file=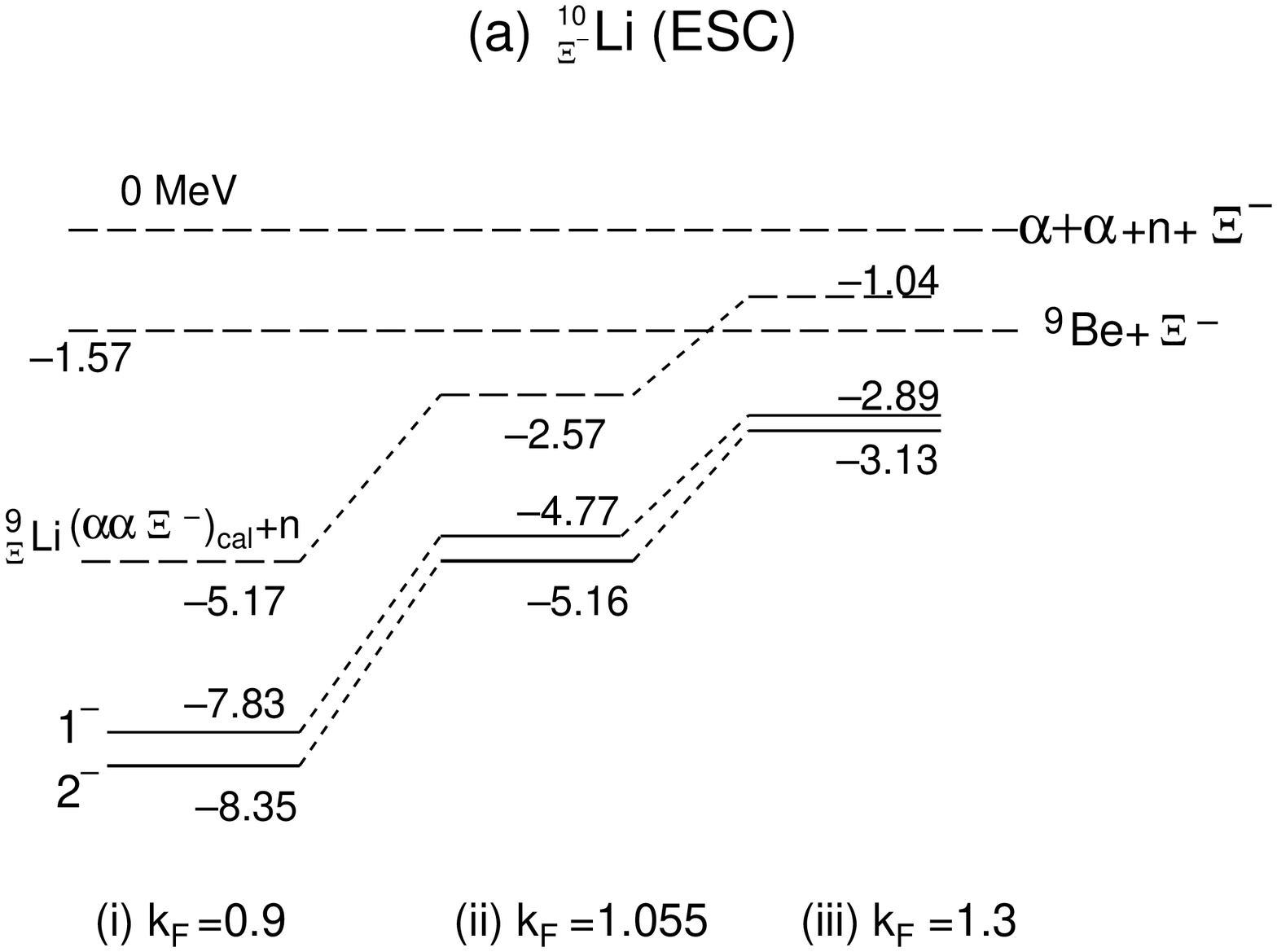,scale=0.40}
\end{minipage}
\hspace{\fill}
\begin{minipage}[t]{8.7 cm}
\epsfig{file=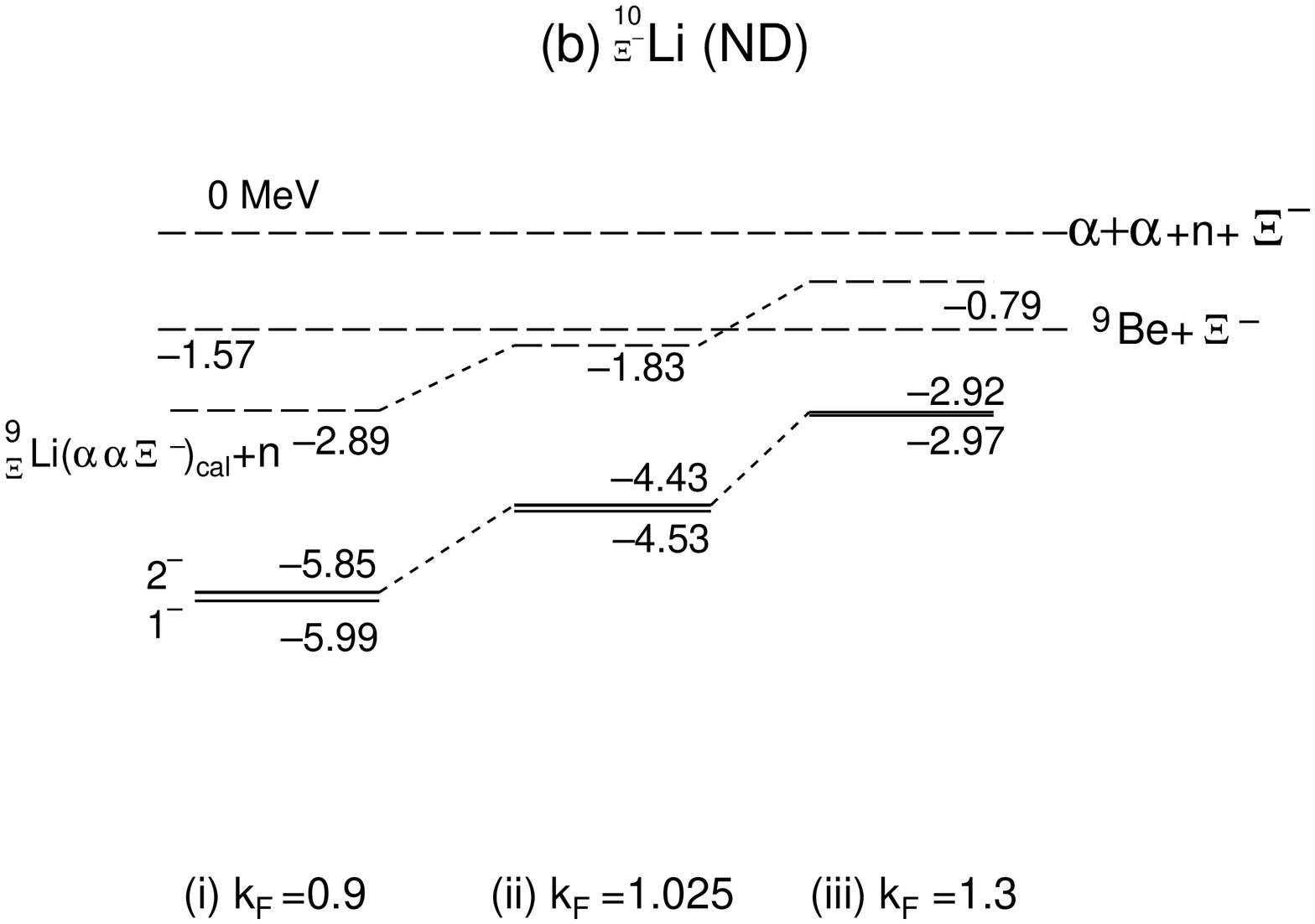,scale=0.40}
\end{minipage}
\caption{(a)Calculated energy levels of
$^{\: 10}_{\Xi^-}$Li for three $k_{\rm F}$ values
 using ESC.
(b)Calculated energy levels of
$^{\: 10}_{\Xi^-}$Li for
three $k_{\rm F}$ values
 using ND.
The energies are shown when the imaginary part of the 
$\alpha \Xi^-$ interaction is switched off.
The energies are measured from the $\alpha +\alpha  +n+\Xi^-$ threshold.
The dashed lines indicate thresholds.}
\label{fig:b10xilevel-esc}
\end{figure*}

%
%===============================================================
\subsection{Results for $^{\: 10}_{\Xi}$Li($\alpha \alpha n \Xi^-$)}

The calculated results for 
$^{\: 10}_{\Xi^-}$Li($\alpha \alpha  n \Xi^-$)
within the four-body model 
are displayed in Fig.~\ref{fig:b10xilevel-esc}
for the lowest $T=1$ doublet state energies
 $(J^\pi=2^-, 1^-)$.
The $3/2^-$ ground state of the core nucleus $^9$Be is bound by 
about $1.57$ MeV with respect to the $\alpha +\alpha +n$ threshold.
We emphasize that, if the $\alpha \alpha \Xi^-$ system is bound 
as shown in Sec.IV, then the
$^{\:10}_{\Xi^-}$Li($\alpha \alpha  n \Xi^-$) 
system is surely expected to be bound  because
the interaction between the $\Xi^-$ and a $p$-orbit neutron
is weakly attractive.

\begin{table}[htbp]
  \caption{The calculated binding energies, $E$
    of
  the $1^-_1$ and $2^-_1$ states in the
  $^{\: 10}_{\Xi^-}$Li($\alpha \alpha n \Xi^-$) system
  for several values of 
  $k_{\rm F}$. The values in parentheses are energies when
  the imaginary part of the $\alpha \Xi^-$ interactions are switched off.
   The energies are measured from the
  $\alpha +\alpha +n+\Xi^- $ threshold.
   The calculated r.m.s. radii, $\bar{r}_{\alpha {\rm -} \Xi^-}$,
  $\bar{r}_{\alpha {\rm -} n}$ and $\bar{r}_{\alpha {\rm -} \alpha}$
   of $2^-$ state using ESC and ND. }
  \label{tab:b10xi}
\vspace{0.2cm}
\hspace{1.0cm} (a) $^{10}_{\Xi^-}$Li(ESC)
\begin{center}
  \begin{tabular}{cccccc}
\hline
\hline
with Coulomb &$k_{\rm F}$(fm$^{-1}$)
&$0.9$  &$1.055$  
&$1.30$  \\
$2^-$ &$E$ (MeV) &$-7.99$ &$-4.83$ &$-2.87$ \\
& &$(-8.35)$  &$(-5.16)$  &$(-3.13)$ \\
 &$\Gamma$(MeV)  &$5.87$  &$3.63$ &$1.71$  \\
&$\bar{r}_{\alpha {\rm -} \Xi^-}$(fm)  &3.05
&3.72 &5.03 \\
&$\bar{r}_{\alpha {\rm -} n}$(fm) &3.55 &3.70 &3.83 \\ 
&$\bar{r}_{\alpha {\rm -} \alpha}$(fm) &3.25 &3.41 &3.54 \\ 
\hline
$1^-$ &$E$ (MeV) &$-7.48$ &$-4.42$  
&$-2.64$ \\
& &$(-7.84)$  &$(-4.77)$  &$(-2.89)$  \\
&$\Gamma$(MeV)  &5.72   &3.44  &1.50  \\
\hline
without Coulomb &$E$ (MeV) &$-5.54$ &$-2.76$ &$-1.41$ \\
$2^-$ & &$(-5.93)$  &$(-3.14)$  &$(-1.63)$  \\
&$\Gamma$(MeV)  &$5.39$  &$3.00$ &$1.10$  \\
\hline
\end{tabular}
\end{center}
\vspace{0.2cm}
\hspace{1.0cm} (b) $^{10}_{\Xi^-}$Li(ND)
\begin{center}
  \begin{tabular}{cccccc}
\hline
\hline
with Coulomb &$k_{\rm F}$(fm$^{-1}$)
&$0.9$  &$1.025$  
&$1.3$  \\
$2^-$ &$E$ (MeV) &$-5.83$ &$-4.42$ &$-2.92$\\
& &$(-5.85)$  &$(-4.43)$  &$(-2.92)$  \\
&$\Gamma$(MeV)  &$0.75$  &$0.42$ &$0.10$  \\
&$\bar{r}_{\alpha {\rm -} \Xi^-}$(fm)  &3.55
&4.10 &5.40 \\
&$\bar{r}_{\alpha {\rm -} n}$(fm) &3.64 &3.72 &3.83 \\ 
&$\bar{r}_{\alpha {\rm -} \alpha}$(fm) &3.35 &3.44 &3.54  \\ 
\hline
$1^-$ &$E$ (MeV) &$-5.98$ &$-4.53$  &$-2.97$ \\
& &$(-5.99)$ &$(-4.53)$  &$(-2.97)$  \\
&$\Gamma$(MeV)  &0.77   &0.43  &0.10 \\
\hline
without Coulomb &$E$ (MeV) &$-3.75$ &$-2.60$ &$-1.54$\\
$2^-$ & &$(-3.76)$ &$(-2.61)$  &$(-1.54)$ \\
&$\Gamma$(MeV)  &$0.62$  &$0.32$ &$0.005$  \\
\hline
\end{tabular}
\end{center}
\end{table}

Although the binding energies of
$^{\: 10}_{\Xi^-}$Li ($J^{\pi}=2^-, 1^-$)  are found to be
fairly sensitive to the choice of the
 $k_{\rm F}$ values, especially, in the
 case of ESC, we think the
 results with $k_{\rm F} \sim 1.0$ fm$^{-1}$ are
 most acceptable.
It is interesting to note that
the $^{\:\ 9}_{\Xi^-}$Li$(\alpha \alpha \Xi^-)_{\rm cal}+n$ threshold
comes below the $^9$Be$+\Xi^-$ threshold in most cases.
It is reasonable that the lowest breakup threshold is 
$^{\:\ 9}_{\Xi^-}$Li$(\alpha \alpha \Xi^-)_{\rm cal}+n$,
because the value of $k_F$ in 
the $A=10$ system has to be $\sim$ 1.0 fm$^{-1}$, 
similar to  the $^{\:12}_{\Xi^-}$Be system.
It is notable that 
the binding energies of $^{\: 10}_{\Xi^-}$Li measured from 
the $^{\:\ 9}_{\Xi^-}$Li$(\alpha \alpha \Xi^-)_{\rm cal}+n$
thresholds 
are similar to each other in both cases of ESC(2.6 MeV)
 and ND (2.7 MeV).

We expect such structure, a valence neutron coupled to
$^{\:\ 9}_{\Xi^-}$Li hypernucleus, 
since the lowest threshold is 
$^{\:\ 9}_{\Xi^-}$Li$(\alpha \alpha \Xi^-)_{\rm cal}+n$.
While, if the $k_{\rm F}$ value
becomes by chance much larger, 
then the $\Xi^-$ particle is coupled to
the ground state of $^9$Be.
Because the lowest threshold is
$^9{\rm Be}+\Xi^-$ [See case (iii) in 
Fig.~\ref{fig:b10xilevel-esc}].

The $2^-$ ($1^-$) state is dominated by the $^{33}S_1$ ($^{31}S_0)$
component of the two-body $n \Xi^-$ interaction.
As demonstrated in Fig.~\ref{Volint},
the $^{33}S_1$ interaction for ESC (ND) is more (less) attractive
than the $^{31}S_0$ interaction.
Therefore, the $2^-$ ($1^-$) state of $^{\: 10}_{\Xi^-}$Li becomes 
the ground state in the case of ESC (ND),
as shown in Fig.~\ref{fig:b10xilevel-esc}.

\begin{figure*}[htb]
\begin{minipage}{0.35\linewidth}
\scalebox{0.35}
{\includegraphics{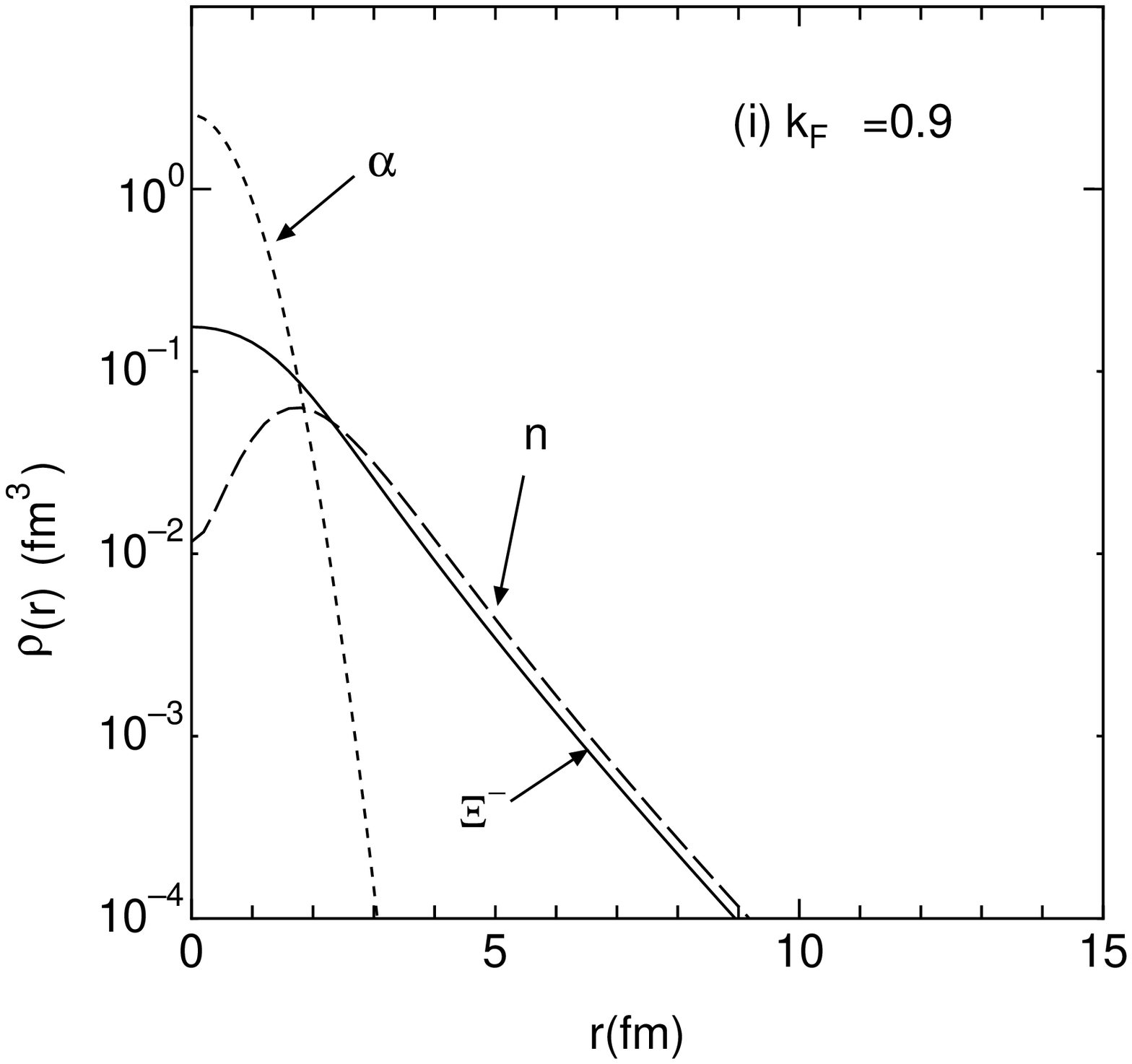}}
\end{minipage}
\begin{minipage}{0.31\linewidth}
(a)$^{10}_{\Xi^-}$Li(ESC)
\vspace{0.5cm}
\scalebox{0.35}
{\includegraphics{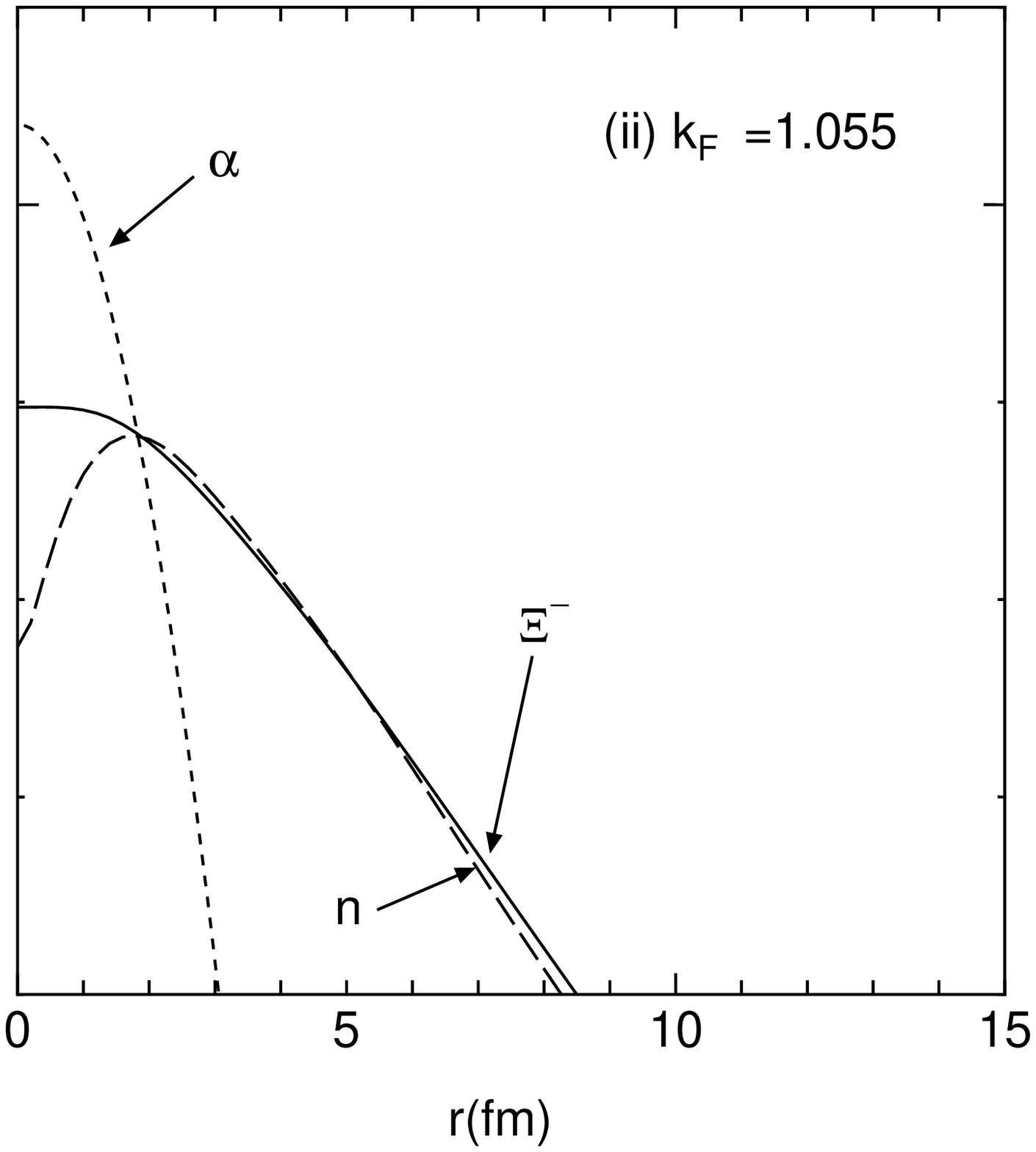}}
\end{minipage}
\begin{minipage}{0.31\linewidth}
\scalebox{0.35}
{\includegraphics{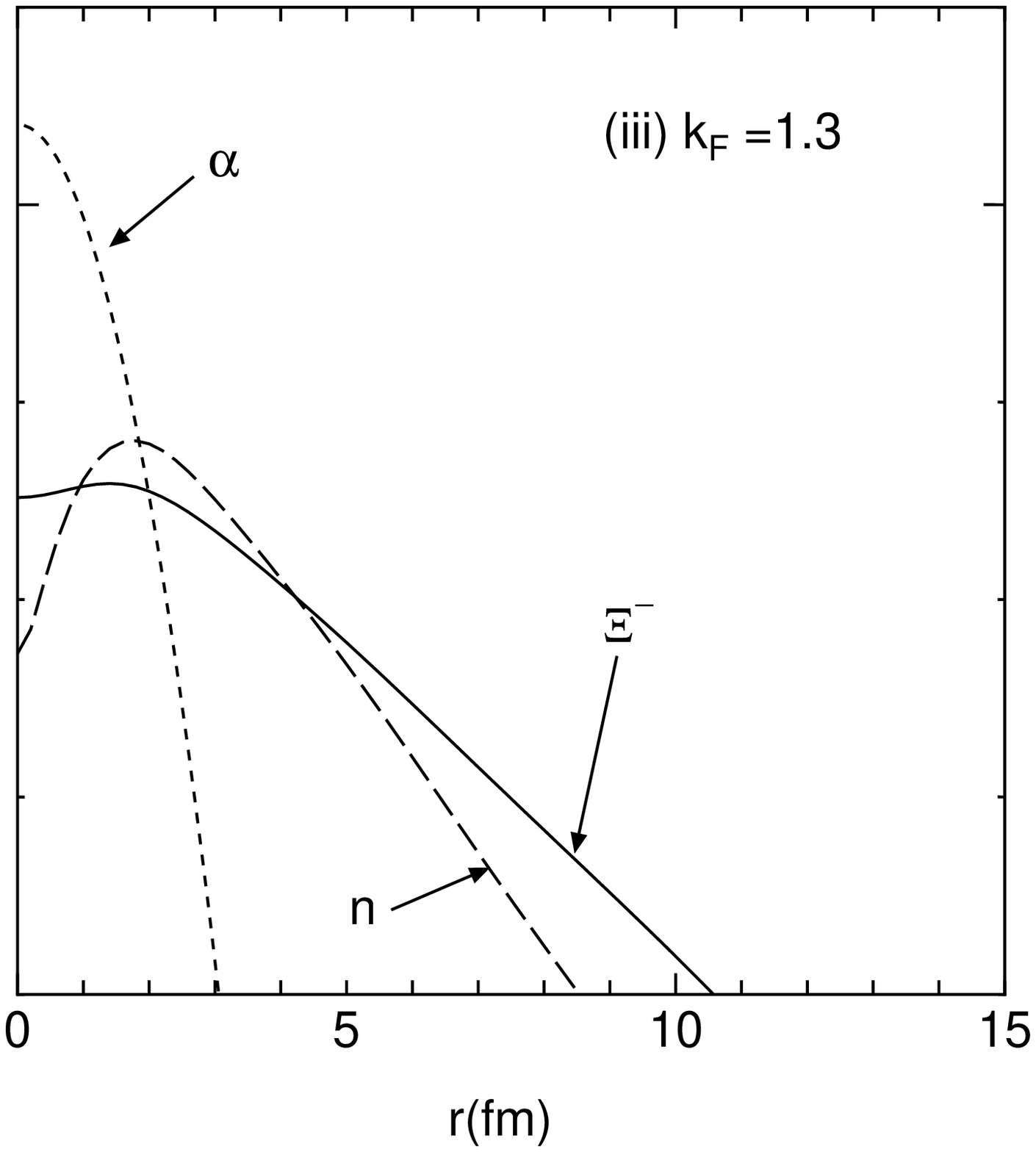}}
\end{minipage}
\begin{minipage}{0.35\linewidth}
\scalebox{0.35}
{\includegraphics{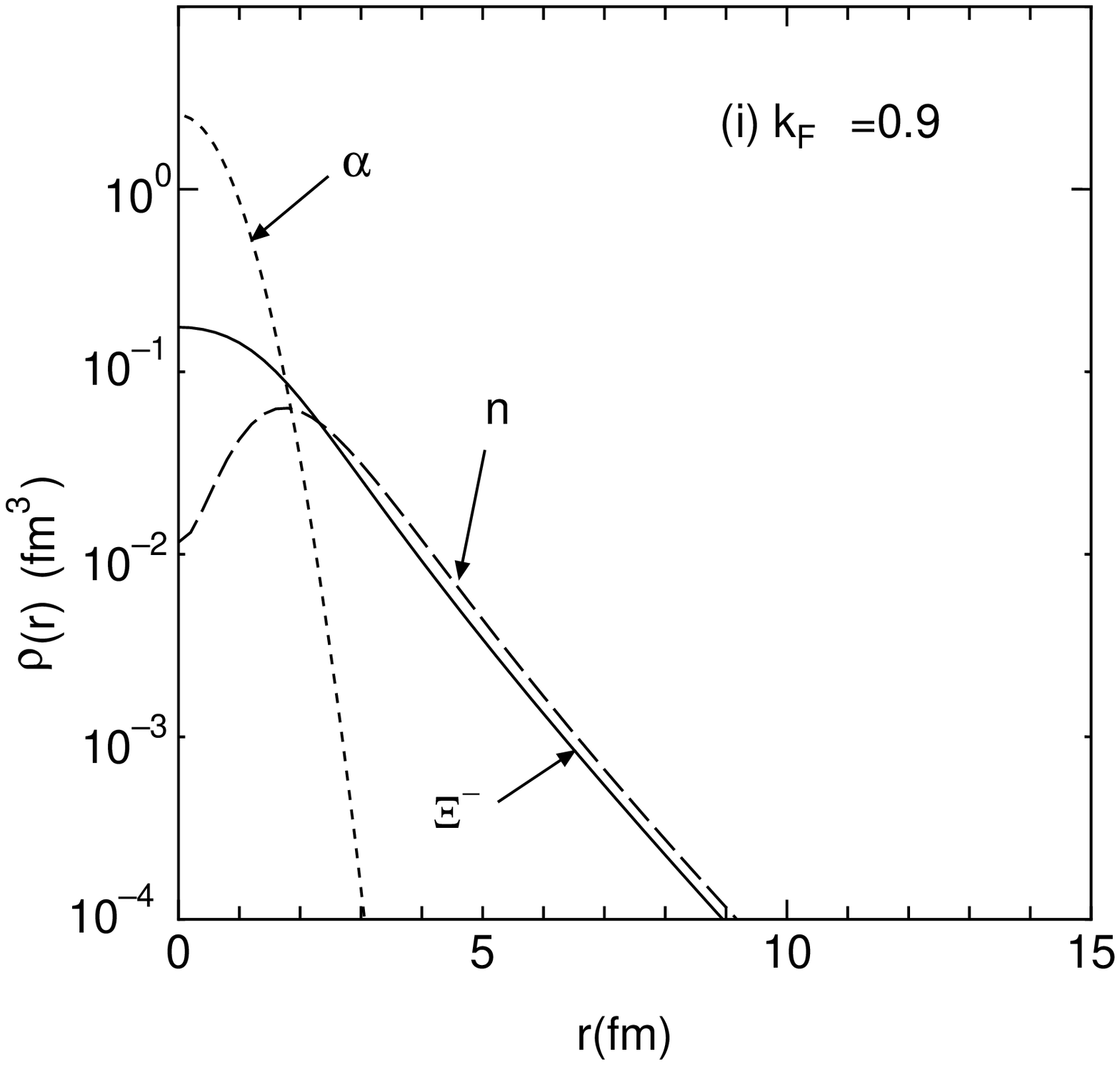}}
\end{minipage}
\begin{minipage}{0.31\linewidth}
(b)$^{10}_{\Xi^-}$Li(ND)
\vspace{0.5cm}
\scalebox{0.35}
{\includegraphics{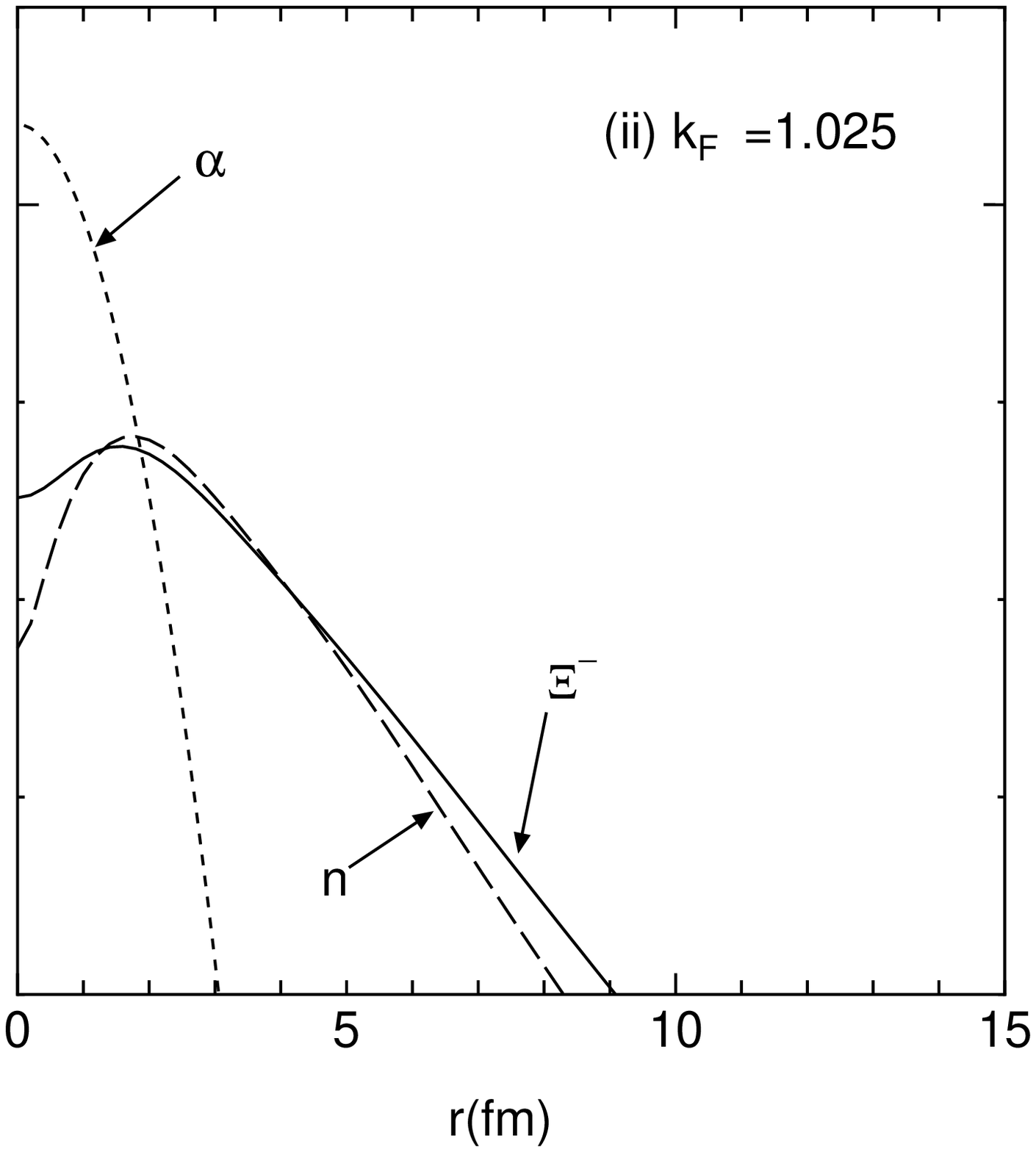}}
\end{minipage}
\begin{minipage}{0.31\linewidth}
\scalebox{0.35}
{\includegraphics{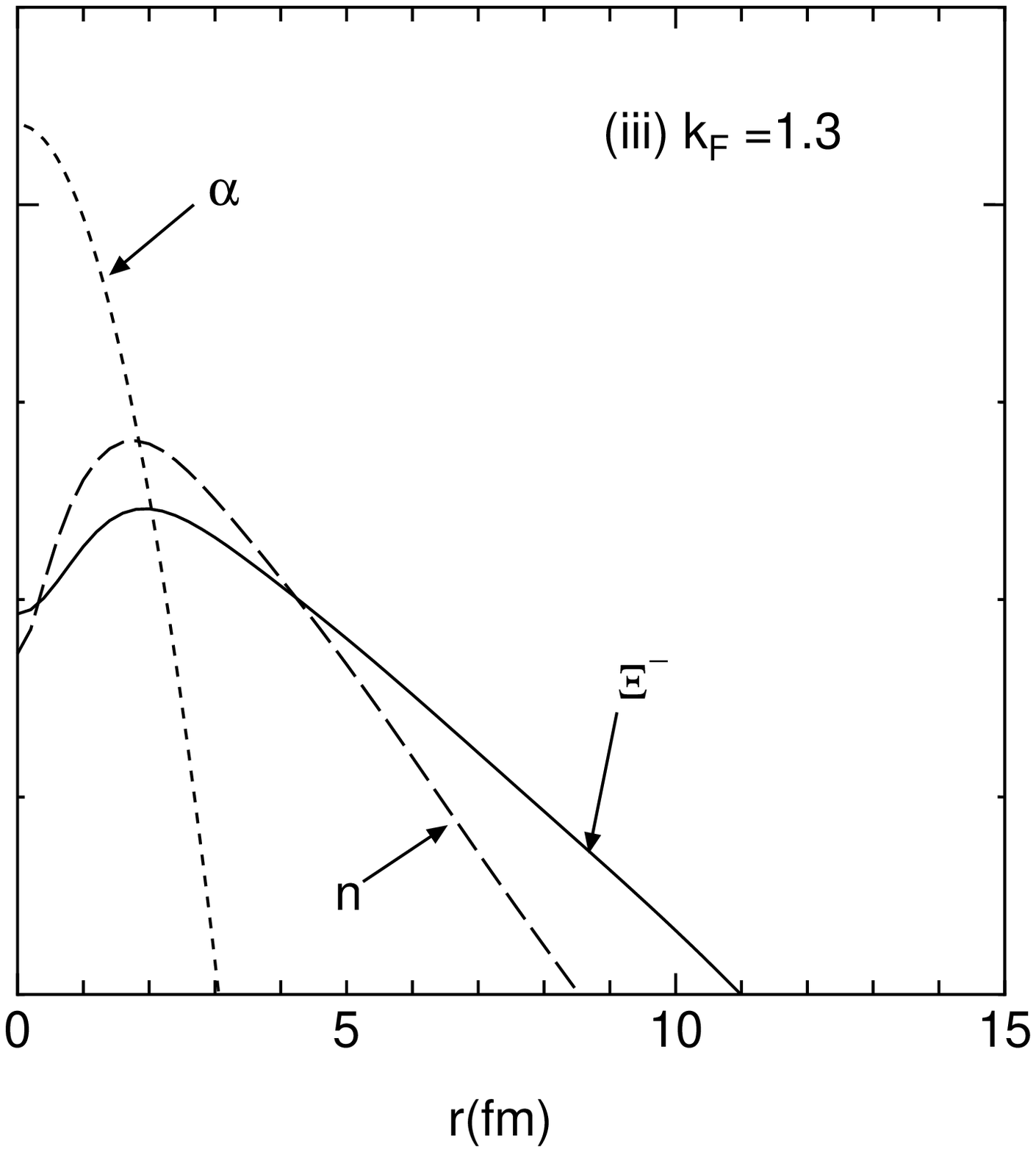}}
\end{minipage}
\caption{(a)Calculated density distribution of
$\alpha$, $\Xi^-$, and a valence neutron
for three  $k_{\rm F}$ values using ESC.
(b)Calculated density distribution of
$\alpha$, $\Xi^-$, and a valence neutron
for three  $k_{\rm F}$ values using ND.
The wavefunctions of $\alpha\Xi^-$ without the imaginary part of the 
$\alpha \Xi^-$ interaction are used.
}
\label{fig:b10xiden-esc}
\end{figure*}

More detailed results are given in Table \ref{tab:b10xi},
which lists also the calculated values of 
the conversion widths $\Gamma$ and the r.m.s. radii, 
${\bar r}_{\alpha {\rm -} \Xi^-}$ and  ${\bar r}_{\alpha {\rm -} n}$.
We show here the results 
with and without the $\alpha \Xi^-$ Coulomb interaction.

As seen in  Table \ref{tab:b10xi},
the decay widths $\Gamma$ calculated
with ESC are much larger than those for ND,
mainly because the 
$^{11}S_0$ $\Xi N$-$\Lambda \Lambda$ coupling interaction
in ESC is far stronger than that in ND.

The r.m.s. distance, ${\bar r}_{\alpha {\rm -} \Xi^-}$,
both for ESC and ND,
are comparable to the 
${\bar r}_{\alpha {\rm -} n}$ values in the cases of choosing
plausible $k_F$ values.
In order to illustrate this situation visually,
we show the density distributions of
$\Xi^-$ (solid curves) and a valence neutron (dashed curves) 
in Fig.~\ref{fig:b10xiden-esc} (a) and (b), where
a single-nucleon density in the $\alpha$ core is indicated by
dotted curves. The upper part (a) and the lower part (b) 
in the figure are for ESC and ND, respectively.
In the $(K^-,K^+)$ reaction,
if the spin-nonflip transition dominates,
the $2^-$ state of $^{\: 10}_{\Xi^-}$Li is selectively excited.
Then, we show the density distributions of $2^-$ states.
The densities of the $\Xi^-$ and a valence neutron
are extended significantly away from the $\alpha$ core.

Let us see the effect of the Coulomb interaction
between the $\alpha$ and $\Xi^-$.
In Table \ref{tab:b10xi}, we list
the binding energy of the $2^-$ state in $^{\: 10}_{\Xi^-}$Li
with and without the Coulomb interaction.
The most important result is that
the states obtained for $^{\: 10}_{\Xi^-}$Li
are bound without Coulomb interactions, namely
as nuclear-bound states,
for all $k_{\rm F}$ values for both ESC and ND.

According to the above calculations,
we can surely expect the existence of the
nuclear bound state with the
predicted $\Xi^-$ binding energies of
$B_{\Xi^-}=3.26$ MeV (ESC, $k_{\rm F}=1.055$ fm$^{-1}$) and
2.85 MeV (ND, $k_{\rm F}=1.025$fm$^{-1}$),
when the imarginary part of $\alpha \Xi^-$ interaction
is taken account of.
These $B_{\Xi^-}$ values seem to be a little
smaller in comparison with the empirical value $\sim 4.5$ MeV
suggested in the $^{12}{\rm C}(K^-,K^+) ^{\:12}_{\Xi^-}$Be
reaction. However, to produce $^{\:10}_{\Xi^-}$Li,
we propose to perform the
$^{10}{\rm B}(K^-,K^+)$ reaction experiment at
J-PARC in addition to that with a
$^{12}$C target.

We say that the $\alpha \alpha  n \Xi^-(^{\: 10}_{\Xi^-}$Li)
 system produced by the
$(K^-,K^+)$ reaction on $^{10}$B is suitable to investigate
$\alpha \Xi^-$ interactions, namely the spin-independent terms
of even and odd-state $\Xi N$ interactions.

%===============================================================
\section{Summary and Outlook}
%===============================================================

In anticipation of priority experiments to be done
at the J-PARC facility, we have carried out detailed
structure calculations for several light p-shell
$\Xi$-hypernuclei, $^{\: 12}_{\Xi^-}$Be, $^{\:\ 5}_{\Xi^-}$H,
$^{\:\ 9}_{\Xi^-}$Li, $^{\:\ 7}_{\Xi^-}$H and $^{\: 10}_{\Xi^-}$Li,
in order to investigate whether we can expect
the existence of bound states of the $\Xi^-$ hyperon.
The calculational framework is microscopic
three- and four-body cluster models using the Gaussian
Expansion Method which has been proved to work quite
successfully in obtaining reliable numerical solutions.

One of the essential issues in preparing such detailed
calculations is what kind of $\Xi N$ interactions one
should use, because there are no definitive experimental
data for any $\Xi$-hypernucleus, and also because there
are large uncertainties in the spin and isospin dependence
in the existing $\Xi N$ interaction models. The only
existing experimental indication, from the
$^{12}$C$(K^-,K^+)^{\: 12}_{\Xi^-}$Be reaction spectrum, is
that the $^{11}$B-$\Xi^-$ interaction is substantially
attractive. However this
constraint is helpful in excluding most of the
$SU_3$-invariant $BB$ interaction models which lead to
repulsive $\Xi$-nucleus potentials.
In this work, we used two  $\Xi N$
potential models, ND and ESC, which give rise to
substantially attractive $\Xi$-nucleus potentials
in accordance with the experimental information.
Although the spin- and isospin-components of these
two models are very different from each other due to
the different meson contributions, we can reliably
speak about the spin- and isospin-averaged properties
such as  $\bar G^{(\pm)}=(G^{(\pm)}_{00}+
3G^{(\pm)}_{01}+3G^{(\pm)}_{10}+9G^{(\pm)}_{11})/16$.
This is why we have focused our attention on the
$\alpha$-cluster based systems and started with an
investigation of the nuclear spin- and
isospin-saturated systems such as
$\alpha \Xi^- (^{\:\ 5}_{\Xi^-}$H) and
$\alpha \alpha \Xi^- (^{\:\ 9}_{\Xi^-}$Li), so as to get a
firm basis of our analyses.

However, the pure $\alpha$-cluster  systems such as 
$\alpha\Xi^-$ and $\alpha\alpha \Xi^-$ cannot be produced
directly, because there are no available nuclear
targets for the $(K^-,K^+)$ reaction. Thus, in order to
explore realistic experimental possibilities, we have extended 
the calculation to the four-body $\Xi^-$-systems having
one or two additional neutrons. This 
explains why we took the $^{\:\ 7}_{\Xi^-}$H$(\alpha nn\Xi^-$)
and $^{\: 10}_{\Xi^-}$Li$(\alpha\alpha n \Xi^-$)
hypernuclei as the typical $\Xi^-$-systems in this
paper.

The major conclusions are summarized as follows:

(1) In order to be consistent with the existing experimental
indication that the $\Xi$-nucleus interaction is attractive, the
fine tuning of the ND and ESC potential models
has been made for applications to $\Xi$-hypernuclei
by adjusting the hard-core radius $r_c$ in ND and the
$\alpha_V$ parameter for the medium-induced effect
in ESC, respectively. Then the $\Xi N$ G-matrices
were derived and represented
in terms of three-range Gaussians with the $k_F$ parameter
expressing its density-dependence within the nucleus.
The $\Lambda \Lambda-\Xi N- \Sigma \Sigma$
coupling term in the bare interaction is
renormalized into the imaginary part in the $G$-matrix
interactions

(2) First we performed the $\alpha\alpha t \Xi^-$
four-body calculation of $^{\: 12}_{\Xi^-}$Be and found
that $k_F=1.055 $fm$^{-1}$ for $G_{ESC}$ and 1.025
fm$^{-1}$ for $G_{ND}$ are most appropriate to produce
the  $B_{\Xi^-}^{\rm CAL}(J=1^-)=2.2$
MeV (without Coulomb interaction) which is that suggested
empirically. These values around $k_F \sim 1.0$ fm$^{-1}$
are  reasonable, because they agree roughly with the
values estimated from the average density in a $A \cong 12$ 
nucleus. Then, we naturally allow smaller $k_F$
values for smaller mass numbers.

(3) In the basic structure calculations for
$\alpha \Xi^- (^{\:\ 5}_{\Xi^-}$H) and
$\alpha \alpha \Xi^- (^{\:\ 9}_{\Xi^-}$Li) systems, we have
tested three values of $k_{\rm F}$ parameters,
$k_{\rm F}=0.9, 1.055$ and $1.3$ for ESC and
$k_{\rm F}=0.9$, 1.025 and 1.3 for ND, respectively.
In the $\alpha \Xi^-$ system, for which
$k_{\rm F}\simeq 0.9$ fm$^{-1}$ is considered to be
reasonable, we obtained only Coulomb-assisted bound
states with small binding energies, since they
disappear without the Coulomb interaction.
In the $\alpha \alpha \Xi^-$ system, on the other hand,
nuclear bound states are obtained for the
acceptable range of $k_{\rm F}$ between 0.9 fm$^{-1}$
and 1.05 fm$^{-1}$. The calculated binding energies
of ESC are larger than those of ND, and also the
$k_{\rm F}$-dependence is more sensitive in ESC.
If these predictions are confirmed, directly or
indirectly, in future experiments, then it will
provide a good check for the spin- and isospin-averaged
$\Xi N$ interaction strengths.

(4) For the lightest realistic example,
$^{\:\ 7}_{\Xi^-}$H$(\alpha  nn\Xi^-$), the four-body
calculation predicts the existence of nuclear
bound states in both cases of ESC and ND at 
reasonable $k_{\rm F}$ values of around 0.9 fm$^{-1}$.
It is interesting to note that the addition of two
neutrons to the $\alpha \Xi^-$ system gives rise
to about 1.3 (2.0) MeV more binding for the ESC (ND)
cases, respectively. If the experiment is carried
out to observe the  $^{\:\ 7}_{\Xi^-}$H bound states, it
is useful to extract information about the even-state
spin- and isospin-averaged part of the $\Xi N$
interaction acting between the $\alpha$ and $\Xi^-$.

(5) For the second realistic example, the
$^{\: 10}_{\Xi^-}$Li($\alpha\alpha n\Xi^-)$ hypernuclus,
we have obtained the nuclear $\Xi^-$ bound
states as a result of careful four-body
calculations with $k_F \sim 1.0$ fm$^{-1}$.
This result is essentially based on the averaged
attractive nature of the $\Xi N$ interactions
acting in the three-body subsystem of
$\alpha\alpha\Xi^-$. It is remarkable to have
similar binding energies of the $J=2^-$ state
for both the ESC and ND interactions
($-4.8$ MeV vs. $-4.4$ MeV with respect to the
$\alpha+\alpha+n+\Xi^-$ threshold). The order of
the doublet states ($J=1^-, 2^-$) is calculated
to be opposite for ESC and ND as a result of the difference
in the $^{31}S_1$ and $^{33}S_0$ components of
ESC and ND acting between the $\Xi^-$ and a neutron.

 In conclusion, it will be quite interesting to
observe the newly predicted bound states in 
future $(K^-,K^+)$ experiments using the $^7$Li
and $^{10}$B targets in addition to the standard
$^{12}$C target. Experimental confirmation
of these states will surely provide us with
definite information on the spin-
and isospin-averaged $\Xi N$ interactions;
note the information on its even-state part
from $\alpha\Xi^-$ and $\alpha\alpha\Xi^-$ and
its odd-state part from
$\alpha\alpha\Xi^-$. Such a plan is a
challenging project in the study of $\Xi$-hypernuclei
that have yet to be explored.
In order to convert the present predictions into
concrete experimental proposals at J-PARC, the
reaction cross sections should be 
estimated for the $^7{\rm Li}(K^-,K^+)^{\:\ 7}_{\Xi^-}$H,
$^{10}{\rm Li}(K^-,K^+)^{\: 10}_{\Xi^-}$Li and
$^{12}{\rm C}(K^-,K^+)^{\: 12}_{\Xi^-}$Be reactions.

%
%-------------  Acknowledgment ----------
%
\section*{Acknowledgments}
The authors thank B.\ F.\ Gibson for helpful discussions.
This work was supported by a Grant-in-Aid for
Scientific Research from Monbukagakusho of Japan.
The numerical calculations were performed on the 
HITACHI SR11000 at KEK.

\end{document}